\documentclass[final,1p,times,twocolumn,authoryear]{elsarticle}

\usepackage{amsmath, amssymb, mathrsfs}
\usepackage{graphicx}
\usepackage{color}

\DeclareGraphicsExtensions{.eps,.ps,.bmp,.tif,.tiff,.tga,.gif} 

\usepackage[framemethod=TikZ]{mdframed}
\mdfdefinestyle{MyFrame}{%
    linecolor=white,
    outerlinewidth=1pt,
    roundcorner=0pt,
    innertopmargin=\baselineskip,
    innerbottommargin=\baselineskip,
    innerrightmargin=0pt,
    innerleftmargin=0pt,
    backgroundcolor=gray}

\usepackage{lineno}

\journal{Journal of Theoretical Biology}

\makeatletter
\def\ps@pprintTitle{%
 \def\@oddhead{\scriptsize{\textsf{Accepted manuscript version for publication in Journal of Theoretical Biology, see http://doi.org/10.1016/j.jtbi.2018.06.019.}}}
 \let\@evenhead\@empty
 \def\@oddfoot{\scriptsize{\textsf{This manuscript version made available under the CC-BY-NC-ND 4.0 license, see http://creativecommons.org/licenses/by-nc-nd/4.0/.}}}%
 \let\@evenfoot\@oddfoot}
\makeatother

\begin{document}
\newcommand{\ba}{{\bf {a}}}
\newcommand{\bn}{{\bf {n}}}
\newcommand{\bx}{{\bf {x}}}
\newcommand{\bu}{{\bf {u}}}
\newcommand{\bv}{{\bf {v}}}
\newcommand{\II}{\mathbb{I}}
\newcommand{\ra}{\rightarrow}
\newcommand{\RR}{\mathbb{R}}
\newcommand{\Rn}{\mathbb{R}^n}
\newcommand{\Sn}{\mathbb{S}^{n-1}}
\newcommand{\Sone}{\mathbb{S}^{1}}
\newcommand{\Stwo}{\mathbb{S}^{2}}
\newcommand{\Bone}{{\mathbb{B}_1(0)}}
\newcommand{\NN}{\mathbb{N}}
\newcommand{\ZZ}{\mathbb{Z}}
\newcommand{\RRn}{\RR^n}
\newcommand{\RRd}{\RR^n}
\newcommand{\IV}{\mathbb{V}}
\newcommand{\BV}{\mathcal{B}(V)}
\newcommand{\BS}{\mathcal{B}(\Sn)}
\newcommand{\M}{\mathcal{M}}
\newcommand{\E}{\mathcal{E}}
\newcommand{\IE}{\E}
\newcommand{\X}{\mathbb{X}}
\newcommand{\A}{\mathcal{A}}
\newcommand{\U}{\mathcal{U}}
\newcommand{\V}{\mathcal{V}}
\newcommand{\ud}{\mathrm{d}}
\newcommand{\pdifft}[1]{\dfrac{\partial #1}{\partial t}}
\newcommand{\T}{\mathscr{T}}
\newcommand{\ep}{\varepsilon}
\newcommand{\Te}{\T_\e}
\newcommand{\pskip}{\medskip\noindent}
\newcommand{\RA}{\mbox{RA}}
\newcommand{\FA}{\mbox{FA}}
\newtheorem{theorem}{Theorem}
\newtheorem{lemma}{Lemma}
\newtheorem{definition}{Definition}
\newtheorem{proposition}{Proposition}
\newtheorem{corollary}{Corollary}

\bibliographystyle{plainnat}

\begin{frontmatter}

\title{Mathematical models for chemotaxis\footnote{See Section 2!} and their applications in self-organisation phenomena}

\author[author1,author2]{Kevin J. Painter}

\address[author1]{Department of Mathematics \& Maxwell Institute for Mathematical Sciences, Heriot-Watt University, Edinburgh, UK, K.Painter@hw.ac.uk}
\address[author2]{Dipartimento di Scienze Matematiche, Politecnico di Torino, Torino, Italy}

\begin{abstract}
\noindent
Chemotaxis is a fundamental guidance mechanism of cells and organisms, 
responsible for attracting microbes to food, embryonic cells into 
developing tissues, immune cells to infection sites, animals towards 
potential mates, and mathematicians into biology. The Patlak-Keller-Segel (PKS) 
system forms part of the bedrock of mathematical biology, a go-to-choice for 
modellers and analysts alike. For the former it is simple yet
recapitulates numerous phenomena; the latter are attracted to 
these rich dynamics. Here I review the adoption of PKS systems when explaining 
self-organisation processes. I consider their foundation, returning to 
the initial efforts of Patlak and Keller and Segel, and briefly describe their 
patterning properties. Applications of PKS systems are considered in their
diverse areas, including microbiology, development, immunology, cancer, ecology 
and crime. In each case a historical perspective is provided on the 
evidence for chemotactic behaviour, followed by a review of modelling efforts;
a compendium of the models is included as an Appendix. Finally, a
half-serious/half-tongue-in-cheek model is developed to explain how 
cliques form in academia. Assumptions in which scholars alter their research 
line according to available problems leads to clustering of academics and 
the formation of ``hot'' research topics.
\end{abstract}
\end{frontmatter}

\section{Introduction}

\begin{figure}[p!]
\begin{center}
{\includegraphics[width=1.0\textwidth]{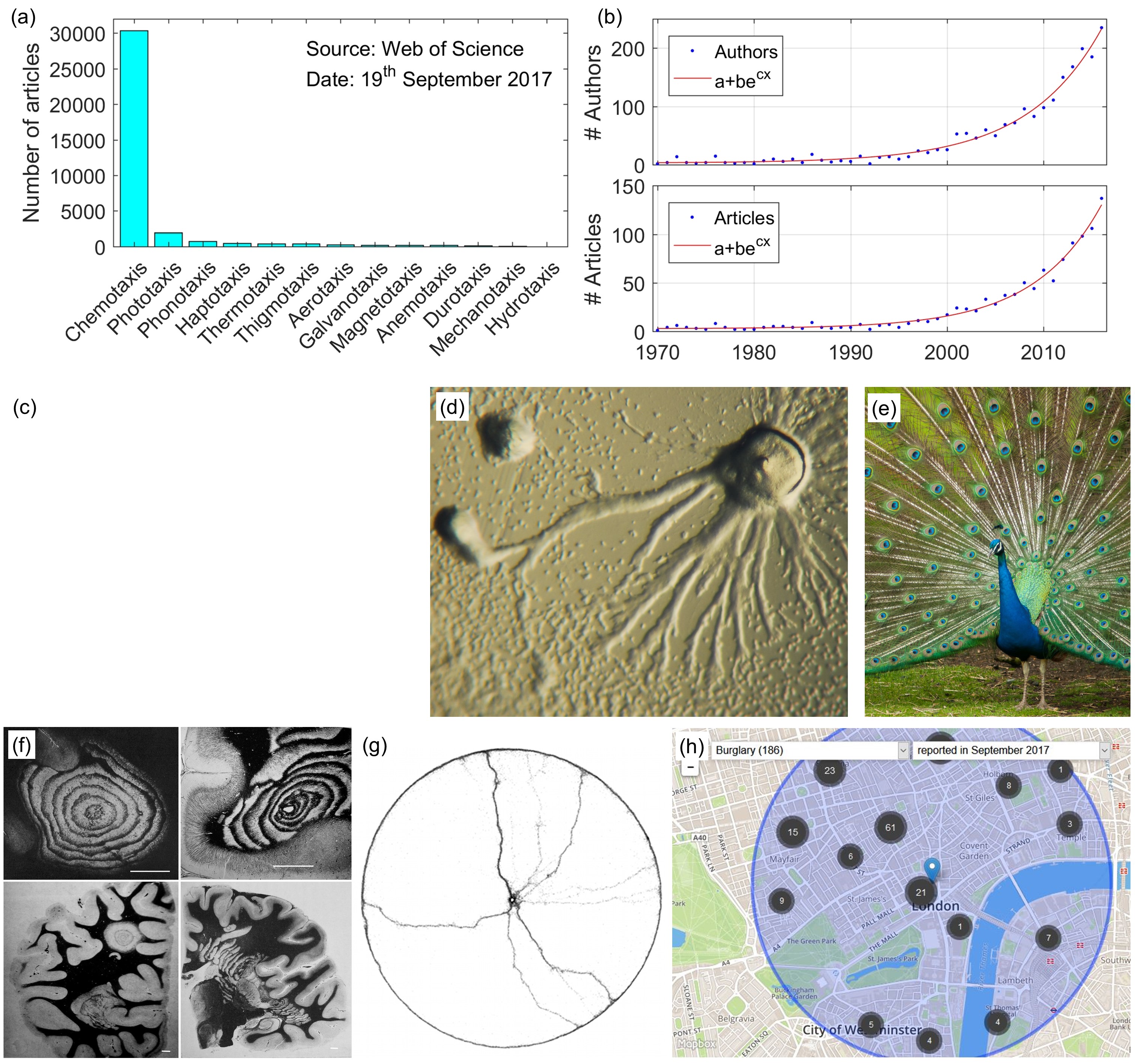}}
\end{center}
\caption{(a) Many ``--taxis'' cues exist, although chemotaxis is by far the most intensely 
studied. The bar chart reflects the number of articles (Web of Science Topic Search, 1900-September 2017) 
for aerotaxis (movement in the direction of oxygen gradients) anemotaxis (wind currents), chemotaxis (chemicals),
galvanotaxis (electric fields), haptotaxis (adhesion gradients), hydrotaxis (moisture), magnetotaxis 
(magnetic fields), mechanotaxis (mechanical cues), phonotaxis (sound), phototaxis (light), rheotaxis 
(water currents), thermotaxis (heat), thigmotaxis (physical contacts). (b) An analysis of the 
number of articles (top) and the number of authors (bottom) citing \cite{keller1970} since 1970. 
Exponential fits provide an excellent match. (c-h) Examples of diverse areas in which PKS models have been applied: (c) Expanding rings of spots in {\em S. typhimurium}, (This figure has been removed due to permission rights); (d) {\em Dd} slime mold aggregation (Figure released into public domain, \texttt{https://commons.wikimedia.org/wiki/File:Dictyostelium\_Aggregation.JPG}); (e) Feather placement during development; (f) Sclerosis patterns in neuropathologies (reproduced from Figure 1 of \cite{khonsari2007}, under the terms of a Creative Commons attribution license); (g) Trail formation in Argentine Ants (reproduced from Figure 1 of \cite{perna2012}, under the terms of a Creative Commons attribution license); (h) Crime hot spot formation (map generated from ``Crime Map'' facility available at \texttt{https://www.police.uk/}).} \label{figure1}
\end{figure}

The ability to detect and migrate in response to guidance cues is 
widespread and multifaceted, transcending scientific boundaries: 
similar mechanisms that orient immune cells towards infections help
animals navigate towards feeding grounds; the friendly mechanisms essential 
for healing our tissues acquire a sinister face when corrupted in cancer 
development. Of the many cues available, movements induced by chemical 
factors (e.g. chemotaxis, chemokinesis) have received overwhelming attention, 
clear reflection of their ubiquitous nature (Figure \ref{figure1}a). As our 
ability to probe the molecular world has increased, phenomena such as 
chemotaxis have emerged as model processes to understand how cells and 
organisms read, respond and shape the sensory information in their environment.

\smallskip
A not insignificant contribution has arisen from mathematical and 
computational modelling. Early studies utilised continuous/population-level 
approaches, via partial differential equation (PDE) systems for the
evolving densities of cells/organisms and the concentrations of attractants/repellents. 
Cheaper computational power has allowed the field to proliferate and diversify
into increasingly sophisticated, specialised forms: for example, detailed 
molecular-level models to describe the complex signalling pathways or agent-based models 
that represent each individual. Consequently, models can be instilled with highly specific 
properties, fostering truly interdisciplinary studies that blend experiment and theory.

\smallskip
Yet, despite the trend towards fine detail, the ongoing development and 
exploration of continuous-level models remains a helpful, higher-level, approach. 
Benefiting from their roots in classic mathematics, they come with  
analytical tools capable of generating insight without recourse to heavy 
number crunching. For chemotaxis models, the system of PDEs 
formulated in \cite{keller1970,keller1971a,keller1971b} and anticipated 
in \cite{patlak1953a,patlak1953b} has proven particularly appealing.
Composed of coupled reaction-diffusion-advection equations, it describes the 
evolving densities of one or more chemotactic population and its 
attractants/repellents, with a natural and logical description 
for the macroscopic {\em consequence} of chemotaxis. Defining 
$u(\bx,t)$ to be the density of the chemotactic population at position 
$\bx \in \RR^n$ and time $t$ and $v(\bx,t)$ as its corresponding chemoattractant, 
the basic Patlak-Keller-Segel (PKS)\footnote{These are often simply stated as the 
Keller-Segel equations in the literature. Here I use PKS to reinforce the 
connection between the distinct approaches employed by Patlak and Keller-Segel 
during early modelling.} model is of the form
\begin{equation}\label{classical}
\begin{array}{rcl}
u_t & = & \nabla \cdot \left( D_u(u,v) \nabla u - u \chi(u,v) \nabla v \right) + f(u,v)\,, \\
v_t & = & D_v \nabla^2 v + g(u,v)\,.
\end{array}
\end{equation}
$f(u,v)$ and $g(u,v)$ respectively describe population and chemoattractant reaction 
kinetics, $D_v$ is the chemoattractant diffusion coefficient and 
$D_u(u,v)$ and $\chi (u,v)$ respectively define population diffusion and 
chemotactic sensitivity coefficients. The key component is the advective {\em taxis-flux} 
choice $u \chi(u,v) \nabla v$: intuitively, this describes population drift up (or down) the 
direction of the local (chemical) gradient. 

\smallskip
These models have readily been developed and applied to problems in fields
ranging from ecology to economics, or cancer to crime, Figure \ref{figure1}c-h. 
In turn this has attracted analysts, imbuing the field with a sophisticated 
(though by no means complete) mathematical underpinning. Year on year 
numbers of publications/researchers citing one of the key early chemotaxis papers are
closely fitted by exponential curves, Figure \ref{figure1}b, 
indicating this line of research continues to grow. Yet, modelling 
demands introspection: Vincent van Gogh said ``Do not quench your inspiration and 
imagination; do not become the slave of your model''. We should not simply stick 
rigidly to a familiar form, blind to its limitations.

\smallskip
In this review, I evaluate the use of PKS models in describing chemotaxis (and other 
taxes), particularly focussing on examples of pattern 
formation/self-organisation\footnote{I apologise to the many authors whose relevant 
work has not been cited for conciseness. In mitigation, a search on ``chemotaxis'' in Web of Science 
generates more than 30,000 studies dating back to 1900...}. I do not review the 
numerous excellent studies that primarily concentrate on their
mathematical analysis: a number of reviews already cover these aspects in depth \citep{horstmann2003,perthame2006,hillen2009,wang2013,bellomo2015}.
Next, the fundamental modelling that
led to (\ref{classical}) is reviewed and an overview given regarding its patterning behaviour. 
I then proceed field-by-field, beginning with its motivations in microbiology 
and sweeping across areas including developmental biology, immunology, cancer, 
ecology and the social sciences. For each case, the historical justification for 
chemotaxis is described and models are discussed; a compendium that contains
many of these models is provided in the Appendix. Finally, I 
demonstrate how PKS models can continue to penetrate new areas, via a novel 
application to explain clique formation in research.

\section{Navigating the nomenclature}

\begin{figure}[t!]
\begin{center}
{\includegraphics[width=1.0\textwidth]{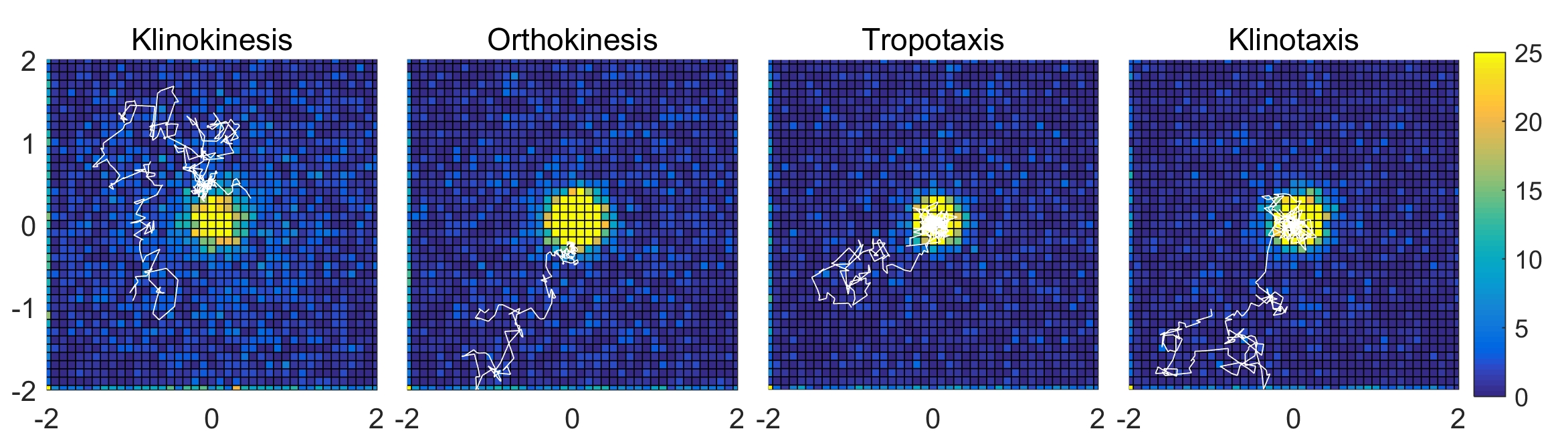}}
\end{center}
\caption{Results of agent-based simulations, where individuals respond to a chemical cue
peaking at the origin (e.g. chemical injected from a micropipette). Agents move via a 
stochastic velocity-jump random walk \citep{othmer1988}, where particles perform: 
(i) klinokinesis, in which turning frequency increases with concentration; (ii) 
orthokinesis, in which speed decreases with concentration; (iii) tropotaxis, in which orientation 
is biased according to an instantaneous chemical gradient; (iv) klinotaxis, in which orientation is 
biased according to a gradient calculated from separate locations at successive time points. 
For each case the path of a single agent (white track) and the histogram (heat map indicates particle number/box) for the particle distribution is plotted, the latter following evolution to a ``steady state'' distribution (10,000 individuals).} \label{figure2}
\end{figure}

\smallskip
Adopting uniform terminology in reviews of chemotaxis is a 
challenge: a varied nomenclature arises in a vast literature spanning 
microbiology, medicine and mathematics, embryology and ecology. 
Oxford English Dictionary's rather precise definition ``The orientated or 
directional movement of a motile cell or organism in response to a gradient 
of concentration of a particular substance; an instance of this.'', is 
countered by the vaguer ``The movement of a microorganism or cell in 
response to a chemical stimulus'' of Collins\footnote{``Chemotaxis'',  \texttt{https://www.oed.com} and \texttt{https://www.collinsdictionary.com}.}. Yet, stating one 
correct definition ignores the historical perspective 
and vagaries of distinct fields. 

\smallskip
Chemotaxis was first described for bacteria and other single cells more than 
a century ago, following pioneering studies of \cite{engelmann1881a,engelmann1881b,engelmann1883} and \cite{pfeffer1884}.
Attributing this to bacteria {\em steering towards} the chemical signal -- 
an implied {\em directed} migration -- the latter coined the term chemotaxis. 
This steering hypothesis was later abandoned by Pfeffer, but 
{\em -taxis} had entered common parlance. In the following years a variety 
of suffixes ({\em -taxes}, {\em -kineses} 
and {\em {-tropisms}}) were attached to movement and/or growth processes, but, as studies surged and distinctions noted, a more formal 
classification was demanded. K{\"{u}}hn's \citep{kuhn1919} 
early systematic classification was superseded by that of 
\cite{fraenkel1940}, and the latter has 
(by and large) stuck for animals and certain cells, such as 
leukocytes \citep[see][]{keller1977}. 

\smallskip
A {\em {-kinesis}} is an {\em {undirected}} movement response, i.e. 
there is no orientation according to the stimulus. Kineses can be subclassified into {\em {orthokinesis}} if the intensity of the signal triggers a change in the speed (or frequency) of locomotion, or 
{\em {klinokinesis}} if the intensity alters the rate of turning. A {\em {-taxis}} forms a {\em {directed}} movement response, such that the 
cell/organism orients with respect to the signal. The response 
is positive or negative if the orientation is towards or away from 
the source. Taxes subclassify into {\em tropotaxis} if the individual orients by directly measuring a 
spatial gradient, or {\em {klinotaxis}} if a gradient is indirectly measured, for example by comparing signal intensities at two different locations and at successive time points.
Different mechanisms place different demands on sensory skills: tropotaxis would require (at least) two, spatially separated, receptors (e.g. two antennae) while klinotaxis requires 
just one, but with an additional ``memory''. Regarding the original 
usage of ``chemotaxis'', the bacteria of Pfeffer turned out to
use a form of klinokinesis rather than taxis. Nevertheless, chemotaxis 
remains the common term in microbiology to describe their behaviour.

\smallskip
While the above classes all describe different {\em microscopic} behaviours, {\em macroscopic} 
outcomes can turn out to be 
similar: for example, all can induce a population to 
accumulate at the source of some chemoattractant if suitable rules are adopted, Figure 
\ref{figure2}. Chemotaxis is often more broadly used  
to describe an observed macroscopic movement 
flux with respect to a chemical gradient, as in the original 
observations of Pfeffer. Given the spectrum of studies covered, I
will often adopt this looser macroscopic interpretation of chemotaxis, 
unless a more specific attribution has been shown.

\section{Continuous models of chemotaxis}

Historical reviews of continuous chemotaxis models must consider the 
landmark works of \cite{patlak1953a,patlak1953b} and
\cite{keller1970,keller1971a,keller1971b}. These studies have piqued interest 
in numerous fields, sparked volumes of analysis and form part of the
infrastructure of spatial movement modelling. They also elegantly 
illustrate distinct modelling approaches.

\subsection{Patlak} 

Patlak's modelling focussed on the individual perspective: 
given a stochastic random walk description for the path traced out by some 
particle, what is the PDE that governs 
the population-level distribution? Karl Pearson had coined the term random walk 
almost half a century earlier, posing his famous drunkard's walk\footnote{The name derives 
from an exchange in the journal Nature between Pearson and Lord Rayleigh, where the 
problem was first proposed.} problem \citep{pearson1905} when modelling 
mosquito population dynamics \citep{pearson1906}, but Patlak strove to extend 
the theory to include non-independence between successive 
steps (so that individuals persisted in a given direction) and the impact from 
external biases. The latter could easily stem from chemotaxis, although the term 
is not even mentioned in 
\cite{patlak1953a}: applications to organism movement, in particular 
klinokinesis, were considered in \cite{patlak1953b}. 

\smallskip
Avoiding details -- an excellent, accessible description of 
Patlak's modelling is found in \cite{turchin1991} -- Patlak derived a modified 
Fokker-Planck equation in the diffusion approximation to his random walk,
\[
u_t = \nabla \cdot \left[ F_1 (\cdot) \nabla \left( F_2 (\cdot) u \right) +F_3 (\cdot, \varepsilon) u \right]\,.
\]
$F_1, F_2$ and $F_3$ are functions explicitly defined in terms of the 
random walk: speeds, run durations and lengths, 
persistence factor and the external bias $\varepsilon$; the latter offers the 
route for including taxis biases. 

\subsection{Keller and Segel}

The initial modelling of Keller and Segel drew inspiration 
from macroscopic phenomena: the suggested driving role of chemotaxis in 
{\em {Dictyostelium discoideum (Dd)}} aggregation (\citealt{keller1970}, see Section \ref{dicty})
and {\em {Escherichia coli (E. coli)}} bacteria band formation (\citealt{keller1971b}, see Section \ref{ecoli}). 
These are phenomena at a population scale, invoking maybe a million cells or 
more, and a classical approach centred on conservation of mass was adopted. 
In its macroscopic sense, (positive) chemotaxis generates a
population drift up concentration gradients and 
accumulation at attractant sources. Hence, a logical 
chemotactic flux is in the direction of the gradient, i.e.
\[
 J_{\mbox{\footnotesize chemotaxis}} = u \chi(u,v) \nabla v \,.
\]
The sensitivity function $\chi(u,v)$ can depend on both 
cell and attractant densities, or even their spatial and/or temporal 
derivatives. Assuming only the above for the flux would state that 
chemotaxis is perfect: movement exactly in the direction 
of the attractant gradient. More realistically, paths deviate 
due to extrinsic/intrinsic stochasticity and
the above is appended with a Fickian diffusion flux. Subsequently 
adding population kinetics and equations for the chemoattractant then 
gives rise to the PKS model (\ref{classical}). 

\subsection{Explicit derivations of PKS models}

The different modelling approaches echo the distinct definitions of 
chemotaxis: phenomenological/mass conservation methods do not address 
individual subtleties, rather they capture chemotaxis in its 
macroscopic spirit; a random-walk derived model can account for movement idiosyncrasy,
for example the focus on klinokinesis in \cite{patlak1953b}.
Despite this, they share essential features: a diffusive 
term stemming from randomness/uncertainty and advection according to 
the direction of external bias. Yet the connection to the PKS model is hinted 
rather than direct, and we mention further works that formally 
establish the link. \cite{alt1980} built on Patlak's work, constructing 
a random walk model whereby runs alternate with 
reorientations and deriving a differential-integral equation for the density 
at position $\bx$, time $t$, moving in direction 
$\theta$ and having started a run at time $\tau$. A PKS 
equation was obtained for the macroscopic population density in certain limits.  
\cite{othmer1988} formally laid out space-jump 
and velocity-jump processes as conceptual random walk models for 
biological motion: in the former, movement occurs through a sequence 
of positional jumps in space (instantaneous transfer between two 
separated points), the latter generalised the descriptions of
Patlak and Alt. In the process an alternative ``mesoscale'' 
continuous model was clarified, the {\em transport equation} that stemmed 
from the velocity-jump process: an evolution equation for a population 
parametrised with respect to position, time and velocity. While more complex
than the PKS model, this has been enthusiastically adopted in many studies, 
including chemotaxis modelling 
(e.g. see \citealt{rivero1989,othmer2002,dolak2005,saragosti2011,pineda2015}), 
blending advantages of a continuous framework and closer connection to the microscopic setting. 

\smallskip
For space-jump processes, \cite{stevens1997} 
explored the continuous equations derived for subtly varying 
chemosensitive movement rules: models were of general PKS form, but with 
differing diffusion/taxis terms that could translate into profoundly 
distinct behaviour. The connection from velocity-jump processes to 
PKS models initiated in \cite{patlak1953a} and \cite{alt1980} 
continued -- see the review by \cite{othmer2013b} -- to 
generate insight into when and if PKS models can provide a reasonable 
approximation. In bacteria with well understood 
signalling and motion closely approximated by a velocity-jump 
process, such as {\em E. coli}, one can even link parameters and functions characterising 
molecular signalling and motor control to the parameters and functions that 
define diffusive/chemotactic sensitivity terms in a PKS model \citep{othmer2013b}. Moreover, 
novel and interesting variations can emerge, such as the ``perpendicular gradient 
following'' that arises from swimming biases \citep{xue2009}, fractional operator 
terms due to non-Poisson type turning rate distributions \citep{estrada2017} or ``flux-limited'' forms \citep{perthame2018}.
Noteworthy, the above derivations rely on ignoring interactions and 
\cite{stevens2000} is noted for providing the 
first rigorous derivation of a PKS equation for a population of stochastic (weakly)
interacting particles. Further derivations of PKS models from stochastic 
models include those of \citet{newman2004,alber2007,chavanis2010}. Reductions to a 
simple PKS model have also been made from more detailed
continuous models, such as incorporating receptor binding and transport \citep{sherratt1994} 
or through multiphase modelling techniques 
\citep{byrne2004}.

\subsection{Self-organisation and patterning}

The model (\ref{classical}) is elegant and intricate: superficially 
simple, yet capable of complex dynamical behaviour. Undoubtedly,
its self-organising capacity has sparked the most interest and, indeed,
was the primary question of \cite{keller1970}. The most carefully studied 
PKS model contains autotactic feedback, whereby
a population produces its own attractant. A ``minimal'' model comprises of
constant/linear functional forms in (\ref{classical}):
$D_u, \chi$ constants, $f(u,v) = 0$ and $g(u,v) = \alpha u -\beta v$,
where $\alpha$ and $\beta$ respectively describe the rates of 
attractant production by the population and decay. Linear stability analysis 
applied about the uniform/homogeneous solution generates the
{\it necessary} instability condition
\[
\chi \alpha u_s > \beta D_u\,,
\]
where $u_s$ is the density of the population at uniform equilibrium. 
In essence, if a sufficiently dense population ($u_s$) shows sufficiently strong 
chemotaxis ($\chi$) and produces sufficient attractant ($\alpha$) then the 
stabilising effects of random motion ($D_u$) and chemoattractant decay ($\beta$) are 
overcome. Further conditions are derived for specific domains and 
boundary conditions. 

\begin{figure}[t!]
\begin{center}
{\includegraphics[width=1.0\textwidth]{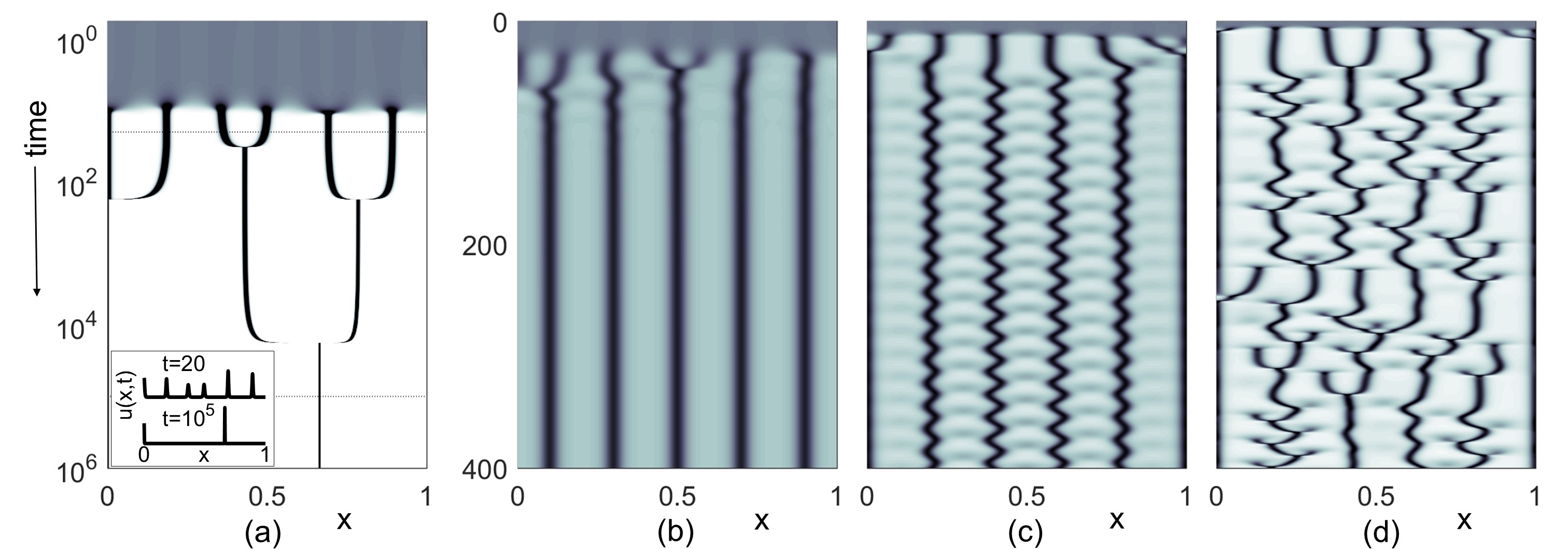}}
\end{center}
\caption{Examples of spatio-temporal patterns in simple 1D chemotaxis models. Grayscale indicates population density from $u=0$ (light gray) to $u \ge 2$ (black). Here $D_u = D_v = 0.001$ and $g(u,v) = u - v$. 
(a) Merging dynamics, for $f(u,v) = 0$ and $\chi = 0.005$. (b-d) Stationary, oscillating 
and chaotic patterns for $f(u,v) = u (1-u)$ and (b) $\chi = 0.005$, (c) $\chi = 0.006$ and
(d) $\chi = 0.008$. The inset of (a) shows the cell density profile at $t=20$ (incipient stage of pattern formation) and $t=10^5$ (following multiple merging events). Simulations solved on a 1D domain $[0,1]$ with zero-flux boundary conditions.}\label{figure3}
\end{figure}

\smallskip
The above intuitively generates symmetry breaking, so what curbs 
the process? Do smooth, stationary density distributions form, do 
they continuously evolve in time or does the population collapse in on itself 
{\it ad infinitum} (``blow-up'')? The answers are non-trivial and explain 
why PKS models have generated so much analytical interest. Other reviews explore 
this subject in depth (most recently \citealt{bellomo2015}, but see also \citealt{horstmann2003,perthame2006,hillen2009}), so here we confine to a 
superficial account. The question of global existence {\em vs} blow-up has 
been extensively studied: in the minimal model solutions exist globally in 1D, 
but for 2D (and above) blow-up occurs if the population exceeds a critical 
density. Blow-up indicates the aggregating tendency, yet is problematic: first,
populations do not form singularities as, even under dense aggregation, 
an individual's finite size (for one) precludes total collapse; second, from 
a numerical/modelling perspective it limits the ability to study dynamics 
post-aggregation. Blow-up can be prevented through {\em appropriate} 
regularisation \citep{hillen2009}, by which we mean density bounds should arise from
natural limiting features: for example, {\em E. coli} aggregates can occur at 
loosely-packed levels \citep{mittal2003}, suggesting that the mechanisms that limit
cell densities are not (solely) the result of volume exclusion.

\smallskip
If patterning occurs, what form? Again the answers are not straightforward. 
As opposed to the large literature on blow-up/global existence, relatively
little is known on the form and stability of solutions to chemotaxis models. For a 1D domain, 
solutions to the minimal model are ``spiky'' \citep{lin1988}, with 
numerical simulations on largish domains generating 
multiple (quasi) regularly-spaced aggregate spikes, each supported by localised attractant 
production. Yet attractant produced by one aggregate diffuses to neighbours, they 
mutually attract and merge until a single ``winner'' remains, see Figure 
\ref{figure3}a. Similar behaviour occurs in related models, such as 
the ``volume-filling'' model, where \cite{potapov2005} have 
numerically explored metastability properties. Essentially, coarsening 
results from transient passage along metastable multi-aggregate steady states
until resting on a stable boundary aggregate; \cite{dolak2005}, 
using singular perturbation methods, showed that neighbouring peaks 
must be sufficiently close to sense each other and merge. Similar properties
for the minimal model have been explored in \cite{hillen2004} and \cite{kang2007}. In the latter
asymptotic methods were used to construct a central spike solution, shown to possess 
metastability and drift exponentially slowly to the boundary. For further results on the
form and stability of patterned solutions to PKS models, see for example
\citet{wang2013,chen2014,zhang2017}. The introduction to \cite{wang2013} is highlighted for
containing a review of earlier results.

\noindent
Obtaining stable multiple aggregate solutions, therefore, seems to demand further 
factors. Adding population kinetics is one method, with classic choices including 
logistic or cubic forms \citep{mimura1996}. The former choice has received 
particular attention and does indeed appear to stabilise multi-aggregate solutions,
but only for certain parameter sets. Others can lead to highly dynamic time-periodic 
or chaotic solutions, where merging alternates with newly forming aggregates \citep{aida2006,wang2007,painter2011,banerjee2012,ei2014}, see 
Figure \ref{figure3}b-d. As illustrated 
later, other models have been shown to generate a variety of further, sometimes bizarre, pattern forms. 
Stable multi-aggregate solutions also appear possible through dual attractant-repellent
systems, where a combination of short-range attractant and long-range repulsion
appears to stabilise multi-aggregate solutions \citep{luca2003}. In short, chemotaxis models 
are capable of wide-ranging dynamical behaviour.

\section{Applications in {\em Dictyostelium} self-organisation} \label{dicty}

\subsection{Background}

Cellular slime molds, the ``social amoebae'', form a celebrated group of 
soil-dwelling microorganisms whose rise to prominence stems from 
a life cycle that straddles individuality and multicellularity. 
The model organism  {\em Dd} \footnote{
First identified in the 1930s by \cite{raper1935} when
searching through camel dung, see \cite{durston2013}.}, 
in particular, serves as a textbook example of self-organisation, see \cite{bonner2009}. 
In its vegetative phase individual cells swarm, consume and divide until 
food is depleted. Starvation heralds a remarkable shift, as the 
dispersed population accumulates into a 
multicellular entity that passes through various stages: differentiating 
and sorting into pre-spore and pre-stalk types, transforming into a 
migrating ``slug'' and eventually a fruiting body. This final act invokes 
apparent altruism, with spore cells shaped into a ball, encased and suspended 
by a thin trunk of sacrificed stalk cells\footnote{In fact, {\em Dd} are not 
so magnanimous: ``cheating'' is rife and certain clones will contribute less 
than their share to the stalk \citep{strassmann2000}.}. Transport of spores to a 
new location restarts vegetative growth. 

\smallskip
Cells collect over an ``aggregation territory'', maybe up to a centimetre 
or so across, with inward movement not smooth but in a sequence of 
pulses that initiate centrally and radiate outwards \citep{arndt1937}.
Closing on the aggregate, cells condense into streams that form a 
branched network, see Figure \ref{figure1}d. A role for chemotaxis was proposed
in the 1940s \citep{runyon1942} and gained substance with the trailblazing 
work of Bonner\footnote{Bonner's videos (available on Youtube) were highly popular, 
with famous visitors to the lab including Albert Einstein.} \citep{bonner1947}: specifically,
a diffusing attractant, generically termed acrasin, was suggested to guide 
cells into a developing mound.
\cite{shaffer1953,shaffer1956} verified that mounds
produced both acrasin and an ``acrasinase'' that degraded it,
leading to a proposed relay system \citep{shaffer1957,gerisch1968,cohen1971a,cohen1971b}: 
(i) central pacemaker cells periodically release an 
acrasin pulse; (ii) nearby cells respond by both moving in its direction 
and releasing further pulses; (iii) the next cell is stimulated 
and so forth. Acrasinase wipes the slate clean, clearing excess extracellular 
acrasin in advance of the next wave. In essence, {\em Dd} forms an excitable medium \citep{durston1973}.
Molecular identities -- acrasin was found to be 3'-5'-cyclic adenosine monophosphate 
(cAMP) \citep{konijn1968} and acrasinase a type of phosphodiesterase 
\citep{chang1968} -- and increasingly sophisticated observations verified the 
relay theory \citep{alcantara1974,shaffer1975,gross1976}: for one of the first times, 
cell signalling was dissected into its raw ingredients and mechanisms. Subsequent decades 
have generated an increasingly nuanced understanding of {\em Dd} chemotaxis, 
see the recent reviews \citet{nichols2015,breitschneider2016}.

\smallskip
As eukaryotes, {\em Dd} cells move similarly to many 
human cells: frequent membrane extensions (pseudopods) form that 
transiently anchor to the substrate/other cells and generate 
traction for motion. In the absence of cAMP, pseudopods extend in random 
directions, but localised detection leads to localised extension. Cells can
directly detect a spatial chemoattractant gradient, possibly as low
as 2\% front to back, and internal signalling amplifies
the initial gradient to create a clear directional polarity \citep{van2004}.
Besides the aggregation of dispersed cells, chemotaxis also directs
other aspects of the life-cycle: pre-aggregation, swarming vegetative cells exhibit 
chemotaxis to folates (secreted by certain bacteria prey) \citep{pan1975}; 
post-aggregation cAMP chemotaxis plays a critical role in mound and slug 
dynamics, with differential chemotaxis sorting pre-spore and 
pre-stalk cells \citep{matsukuma1979,sternfeld1981,traynor1992} and cAMP waves continue to propagate through 
the mound and slug \citep{dormann2001}. 

\subsection{Modelling}

The above process generated source inspiration in the early chemotaxis modelling of
\cite{keller1970}. At the time, chemotaxis/acrasin interactions were known to be 
necessary, but were they sufficient? Specialised ``founder cells'' 
were speculated to act as mound initiators (e.g. \citealt{shaffer1961}), 
but were they essential? Could aggregation occur without cell heterogeneity? 
Modelling allows hypotheses to be stripped to their bare essentials 
to address such questions. Keller and Segel's model (see (\ref{A1}) in 
the model compendium, Section \ref{compendium}) consisted of 
4 variables: amoebae, and concentrations of acrasin, 
acrisinase\footnote{Although formally identified at the time, this was 
recent and generic designations were retained in the model description.} and the
complex formed following their reaction. Pseudo-steady state
approximations allowed further reduction to a familiar two variable 
model for amoebae and their attractant, (\ref{A2}). A Turing-type \citep{turing1952} 
stability analysis\footnote{Turing's classic theory was relatively 
new at the time.} was subsequently employed to show a 
dispersed population could self-organise, given: (i) sufficient
sensitivity to the attractant; (ii) cells produce sufficient 
acrasin; (iii) acrasinase degrades acrasin sufficiently fast. Bonner's 
lab indeed found that sensitivity to cAMP and its production rate 
(dramatically) increased at the inception of aggregation. 

\smallskip
The model was simple even for its time: in their words, 
based on ``simplest possible assumptions consistent with the 
known facts''. Thus, it did not account for a more sophisticated 
periodic cAMP relay system and was perhaps more appropriate for 
related species (such as {\em Dictyostelium minutum}) that secrete 
cAMP steadily and produce simpler, smoother aggregations. 
Its modesty, though, is the base of its appeal: 
it showed that an entirely homogeneous population could 
organise through simple self-secretion of an attractant, i.e. the 
founders could be the entire population. \cite{nanjundiah1973} 
performed a more detailed analysis, suggesting that it could potentially 
give rise to streaming/branching type phenomena.
Following years witnessed significant modelling, but usually 
targeted at the signalling pathways necessary for the periodic cAMP 
relay waves: a comprehensive review can be found 
in \cite{othmer1998}. H{\"{o}}fer {\em et al}
\citep{hofer1994,hofer1995a,hofer1995b} merged 
more refined signalling with a PKS model for 
chemotactic cell movement (\ref{A3}); spatial cAMP signalling had been studied, 
but for ``immobilised'' cells \citep{tyson1989,monk1990}.
Remarkably, the model replicated many complex 
{\em Dd} aggregation phenomena, from outwardly spiralling waves of cAMP to 
streaming and branching of cells as they approached the aggregate.
Other continuous models have also been developed for 
{\em Dd} development, particularly to explain slug regulation 
and movement \citep{pate1986,odell1986,vasiev2003,pineda2015}, although 
their underlying model framework is somewhat distinct from the PKS 
system.

\smallskip
Concurrently, various groups (e.g. \citealt{van1996,dallon1997,savill1997,palsson2000,maree2001}) were 
formulating ``hybrid'' models: agent-based models 
coupled to continuous equations for chemicals. The cell description varied, 
but models shared a common capacity to incorporate explicit 
microscopic detail and numerous features of {\em Dd} morphogenesis could be 
captured. The modelling shift towards the cellular/microscopic scale
followed the greater biology focus at this level: population-scale modelling of {\em Dd} 
chemotaxis via continuous approaches has somewhat taken a backseat in recent 
years, although see \cite{ferguson2016} for a recent example.

\section{Applications in Bacterial Chemotaxis} \label{ecoli}

\subsection{Background} 

The studies of Engelmann and Pfeffer (see \citealt{berg1975} for a review) 
formed initial steps along the road to our most well understood signalling system: 
bacterial chemotaxis in {\em E. coli}. Swimming {\em E. coli} (and certain other bacteria) 
alternate between ``runs'' and ``tumbles'', powered by the rotating flagella 
attached to their cell surfaces. 
During the former, anti-clockwise rotations bundle the flagella 
to generate almost straight-line swimming; tumbling is induced by clockwise 
rotation, with flagella flaying out so that the cell spins quasi-randomly. 
Their chemotactic properties were elegantly studied by 
Adler in the 1960s \citep{adler1966,adler1969} who, by linking 
extracellular chemoattractants (repellents) to specific cell surface 
receptors \citep{adler1969}, lay the groundwork for discerning how a cell 
perceives its chemical landscape. The chemoreceptor array 
stimulates intracellular signalling, mediated by a family of ``Che'' molecules, 
to modulate flagella rotation: see \citet{wadhams2004,parkinson2015} for reviews.
{\em E. coli} are considered too small to discriminate an actual spatial 
gradient (but see \citealt{thar2003}). Chemotaxis rather occurs via 
``klinokinesis with adaptation'': temporal calculations  
in which the absolute extracellular chemical concentration feeds into the 
tumbling frequency \citep{schnitzer1990}. The interval between tumbles lengthens when 
concentration increases so that, overall, more time is spent 
moving up a gradient than down it and cells accumulate near the attractant
source. Longer timescale adaptation resets the system, 
extending the concentration range for effective chemotaxis. 
Summarising, {\em E. coli} do not perform ``true chemotaxis'' under its formal
microscopic definition.

\smallskip
Chemotactic studies have contributed to a perceptional shift of bacteria as
highly sophisticated, capable of intercommunication 
\citep{shapiro1998}. Self-organisation phenomena abound and multicellular patterns 
are formed of astonishing intricacy, see \citet{benjacob2000,kaiser2003}. 
The experiments of Adler revealed chemotactic group behaviour 
in {\em E. coli}, whereby high density travelling bands (or spreading rings) formed 
following insertion of a population into nutrient containing capillary tubes (or petri-dishes) 
\citep{adler1966}. Chemotactic-driven self-organisation was demonstrated 
by \cite{budrene1991,budrene1995}, using
{\em E. coli} and {\em {S. tymphimurium}} populations. Under certain nutrient 
environments patterns formed in the wake of outward radiating waves, 
including stripes and spots of high bacterial density. 

\subsection{Modelling}

\begin{figure}[t!]
\begin{center}
{\includegraphics[width=0.98\textwidth]{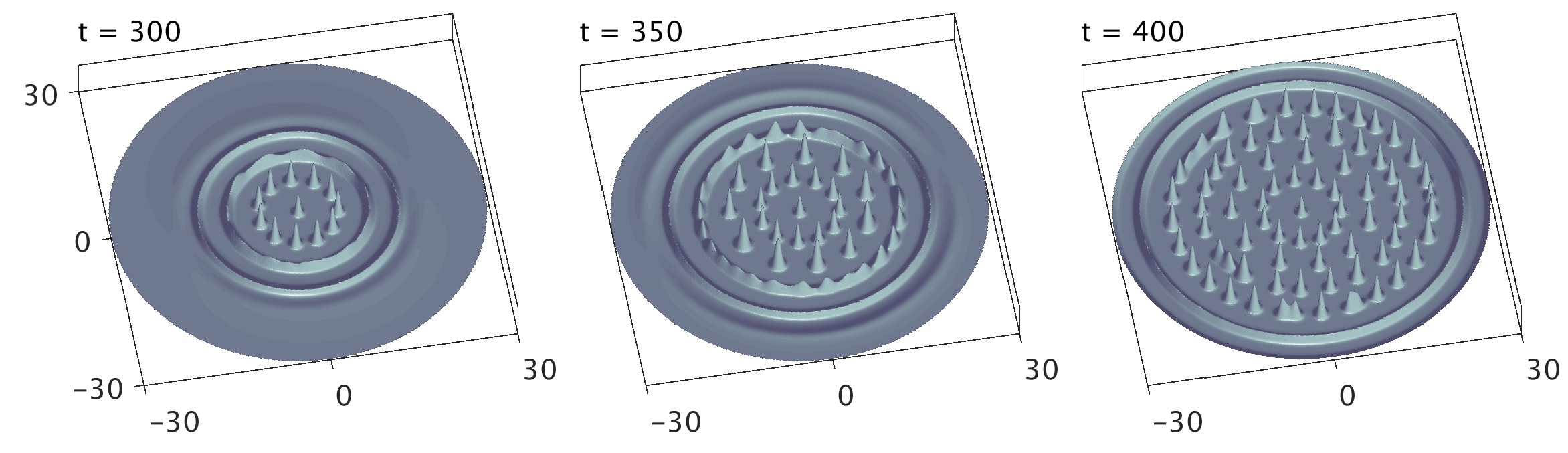}}
\end{center}
\caption{An expanding pattern of spotted rings in the model (\ref{B1}), where we
set $\chi = 3, k_1 = 2, k_2 = 0.03, k_3 = 1, s = 2, k_4 = 0.5, k_5 = 1, k_6 = p = 0, D_u = 0.1$ 
and $D_v = 0.3$. The food source is set at $s=2$ and we solve on a circular domain of radius 30 with zero-flux boundary conditions at the edge. Solutions plot bacteria densities at $t = 300, 350, 400$, following the initial placement of a small compact population at the origin.}\label{woodwardsims}
\end{figure}

Explaining the bacteria bands  observed in Adler's experiments
had been another motivating factor for early modelling \citep{keller1971b}, and
various further studies have explored this phenomenon via PKS models (e.g. \citealt{lapidus1978,lauffenburger1984,brenner1998,croze2011}); we also 
refer to \cite{wang2013} for a review of the mathematics behind travelling 
waves in chemotaxis systems. The use of PKS models to describe the self-organisation 
phenomena observed in \citet{budrene1991,budrene1995} has been a fertile research 
area, see \citet{woodward1995,tsimring1995,tyson1999a,tyson1999b,polezhaev2006,aotani2010}.
In an early application, \cite{woodward1995} integrated experiment 
and theory to determine whether {\em {S. tymphimurium}} patterns could be attributed 
to chemotaxis: experiments reveal spots or solid rings as the 
nutrient (succinate) concentration varied (Figure \ref{figure1}c), while biochemical analyses revealed 
aspartate (a known chemoattractant) secretion by the aggregates. A bare-bones 
PKS model for bacteria and chemoattractant (\ref{B1}) was able to recapitulate
the experimental observations, capturing the pattern transition as the nutrient level
changed; see Figure \ref{woodwardsims} for an illustration. Thus, chemotaxis 
may be a sufficient mechanism for this process of self-organisation. 
Further studies expanded to include additional factors, for example evolving nutrient 
dynamics \citep{tsimring1995,tyson1999a,tyson1999b,polezhaev2006,aotani2010}, 
cell transitions into vegetative form \citep{tsimring1995,polezhaev2006,aotani2010}
and the inclusion of waste products \citep{tsimring1995}: see models (\ref{B2}-\ref{B6}). 
Consequently, many of the even more spectacular observations could 
be captured. An apparently chaotic pattern of {\em E. coli} merging/emerging events has been 
modelled via a PKS system in \cite{baronas2015}.

\noindent
{\em Bacillus subtilis} swimming within fluid environments can create highly 
intriguing dynamics, such as falling plumes and aggregates
at the contact lines of thin fluid layers separating a solid substrate and air
\citep{hillesdon1995,tuval2005}. Bacteria swim upwards through a response to 
oxygen (aerotaxis), which they consume, before
gravitational forces push them downwards. This overall motion results in bioconvection, 
a bacteria-driven fluid flow that convects both the bacteria and oxygen. A
chemotaxis-flow model, whereby a PKS model is coupled to a fluid flow equation (model \ref{B8}),
was developed in \cite{hillesdon1995} and has been readily applied (e.g. \citealt{tuval2005,chertock2012,lee2015,deleuze2016} amongst others) to describe
flow-driven bacteria pattern formation.

\smallskip
Contemporaneous to much of the above, a large literature has emerged 
where the modelling is focussed at the molecular scale: see \cite{tindall2008a,tindall2008b,othmer2013,tu2013} for 
reviews. As for {\em {Dd}}, discrete and/or hybrid approaches have become
increasingly popular for obtaining population-level insights 
while retaining intracellular/cellular scale details: \cite{bray2007,xue2011} give 
two examples of this large literature. Yet, the potentially vast size of colonies 
still preclude their usage for entire populations, which can easily run to billions. Continuous models
therefore remain a valuable tool, but quantitative comparison will increasingly demand 
their construction according to relevant microscopic details.

\section{Applications in embryonic development}

\subsection{Background}

The striking semblance of {\em Dd}'s lifecycle to development was noted in the seminal 1950s 
textbook by \cite{waddington1956}, with speculation that similar mechanisms 
could act. Embryogenesis is, perhaps, the most remarkable direct 
example of self-organisation: a fertilised cell doggedly transforms itself into a 
multicellular organism, orchestrated by proliferation, migration, differentiation, apoptosis, 
intra/extracellular signalling and shaped by internal and external mechanical forces. 
It is an area rife with famous theories, from Turing's reaction-diffusion mechanism 
\citep{turing1952} to pre-pattern models based on position information \citep{wolpert1969}. 

\smallskip
Development is pragmatically studied piecemeal: the transformation of the 
early embryo into a multi-layered tissue structure 
during gastrulation; laying the developmental blueprint during segmentation/somitogenesis; 
within-tissue patterning of individual organs such as the central nervous system, 
skin, lungs and vasculature. {\em Definitive} demonstrations of chemotaxis 
are difficult \citep{shellard2016}: cultivation and study 
{\em in vitro} is possible, but determining whether the same behaviour occurs 
{\em in vivo} is non-trivial. These difficulties aside, a hypothesised role for 
chemotaxis has a history stretching back more than a century. Ram\`{o}n y Cajal\footnote{Father 
of modern neuroscience, and winner of the Nobel Prize in Medicine and Physiology in 1908.} proposed that chemotaxis\footnote{{\em Chemotropism} is more apt, as developing neurons grow ({\em -tropism}) rather than migrate: the 
neuron's ``growth cone'' (a large cytoskeletal cell extension) migrates through the tissue 
and establishes the track of the lengthening axon.} may be key to establishing 
interneuronal connectivity during nervous system development 
\citep{cajal1892}, a theory that lay dormant for decades before spectacular 
revival in the 1990s when various molecular families (e.g. netrins, semaphorins) were
identified with attractive/repulsive guidance information: the review in 
\cite{tessier1996} chronicles the lead-up to these fundamental discoveries, while a 
more recent perspective is provided in \cite{kolodkin2011}. In the 1940s, Twitty proposed that (negative) 
chemotaxis could control dispersal of pigment cells from the neural crest 
\citep{twitty1944,twitty1948,twitty1954} and, while this remains 
unverified \citep{erickson1983}, other neural crest populations can be
chemotactically guided by factors including VEGF \citep{mclennan2010} and 
Sdf1 \citep{theveneau2010} (see also \citealt{shellard2016}). Other
promising cases of chemotaxis during development include border cell migration
\citep{montell2012} and gastrulation  \citep{yang2002}.

\smallskip
The extent to which chemotaxis occurs in the periodic patterning processes beloved by 
mathematical biologists is difficult to deduce. Generation of repeating patterns 
is a common theme -- examples include skin structures 
(pigmentation patterns, hairs, feathers, scales, sensory bristles), tooth 
morphogenesis, taste buds and limb skeletal structures --- yet 
unravelling enough of the complex molecular/cellular/mechanical interactions 
and making firm statements is a challenge. A promising example lies in 
feather/hair arrangements. Feathers emerge from an orderly arrangement of 
skin placodes, each marked by epidermal signalling underlaid by dermal cell
aggregation. In chickens, FGFs \citep{song2004,lin2009,ho2018} and 
BMPs \citep{michon2008} are potential chemotactic factors that may act 
to guide mesenchymal cells into the initial clusters; 
in mammals, FGF-dependent cell clustering occurs prior to hair follicle 
formation \citep{glover2017}. Yet these remain examples of 
chemotaxis under its macroscopic guise: a clustering of 
cells according to a molecular distribution and the precise mechanisms 
involved are uncertain. Additional complications lie in the likely involvement of 
other pattern forming mechanisms, such as those based on activator-inhibitor 
principles (for a review, see \citealt{painter2012}).

\subsection{Modelling}

The mechanochemical framework proposed by Murray, Oster and others 
in the 1980s \citep{murray2003a} offered a general method for 
describing the interactions between motile cells and 
their environment: movement due to force-based 
interactions between cells and the surrounding extracellular 
matrix (ECM) and PKS-type fluxes to describe chemotactic,
haptotactic {\em etc} responses were included. Numerous development-oriented
applications have been made (e.g. see \citealt{murray2003a}), including mesenchymal 
condensation and accumulation of chondrocytes during bone formation 
(chondrogenesis). Of course, most such models do not rely on 
chemotaxis alone: in fact spatial structure can potentially form in the 
absence of active cell movement. Mechanical models with a strong 
chemotactic element have particularly been studied for
({\em in vitro}) vasculogenesis (formation of the capillary network), and
we refer to \cite{ambrosi2005} for a review of this field.

\smallskip
The first detailed discussion of a ``pure'' PKS model in a developmental context 
was put forward for limb chondrogenesis. Limb development is mathematically appealing, featuring a bifurcation process 
in which modal transitions occur as the limb extends: for example, the arm 
bifurcates from the 
single humerus (upper arm) to the radius and ulna (forearm), before 
further bifurcations generate the hand and fingers. \cite{oster1989},
suggested a simple model (\ref{C1}) whereby chondroblasts secreted their own
chemoattractant, discussing its merits with 
respect to generating realistic bifurcation sequences. Further 
analyses by Maini and others \citep{maini1991,myerscough1998} extended to 
a quasi-2D cross section in which new bone is laid, explored heterogeneous 
steady states and their bifurcations and addressed the role of boundary 
conditions and tissue growth. Chemotaxis is certainly a plausible component of the 
chondrogenesis process -- e.g. \cite{mishima2008} -- yet unsurprisingly it 
is significantly more complex: in fact, various studies suggest it may act as a
melting-pot of various classic theories (e.g. positional information and Turing/RD 
ideas), see \cite{green2015}. An obvious challenge for theoreticians is assessing 
the precise contributions of the various pattern generating mechanisms.

\smallskip
PKS models have been proposed in various other developmental processes, 
including pigmentation patterning of snakes \citep{murray1991} and 
fishes \citep{painter1999}, gastrulation \citep{painter2000}, neural crest invasion 
\citep{landman2003,simpson2006}, and feather morphogenesis 
\citep{michon2008,lin2009,painter2018}. While the suggestion of chemotaxis 
was typically speculative at the time, it has since been placed on 
firmer biological foundations: e.g. during gastrulation \citep{yang2002} and 
neural crest invasion \citep{shellard2016}. Applications to feather formation, in 
particular, illustrate a closing gap between modelling and experiment with 
a number of integrated experimental/theoretical studies that utilise the PKS framework
\citep{michon2008,lin2009,painter2018,ho2018}. The model in \cite{lin2009}
was of standard PKS form (\ref{C1}), featuring a population of chemotactic mesenchymal
cells that regulate production of their attractant to generate the dermal cell
aggregates at future feather sites. More detailed models have been proposed in 
\cite{michon2008}, where mesenchymal cells interact with a molecular network 
capable of pattern formation through Turing/activator-inhibitor instability (\ref{C2}),
and in \cite{painter2018}, where chemotactic-driven aggregation is generated 
mediated by epithelium signalling and molecular (FGF and BMP) regulators, 
model (\ref{C3}). The capacity of these models to replicate feather placode
patterning and predict experimental perturbations offers a tantalising 
prospect for targeted future modelling.

\section{Applications in physiology and disease}

\subsection{Background}

Adult tissues and organs demand regeneration, monitoring 
and repair, invoking cell migration and chemotaxis amongst the 
numerous processes. The scope of this section limits us to a
superficial coverage and we restrict to a 
few works of historical note. Reports of chemotactic responses in 
the immune system date to the late 19th century observations of 
\cite{leber1888} and \cite{metchnikoff1892}\footnote{Regarded as the
founder of cellular immunology, and awarded the Nobel Prize in 
Medicine and Physiology in 1908.},  
where phagocyte accumulation at infection sites was hypothesised to 
result from chemotactic guidance. Decades later, development of an 
{\em in vitro} assay in \cite{boyden1962} (the ``Boyden chamber'') allowed
(immune cell) chemotaxis to be studied quantitatively and 
reproducibly and a large family of {\em chemokines} 
({\em chemo}tactic cyto{\em kines}, \citealt{griffith2014}) are now known to 
control physiological chemotaxis, not just for immune system responses but 
many other processes. For example, during wound healing an assortment of
chemokines act to guide cells from the early inflammation stage to 
final remodelling \citep{gillitzer2001}.

\smallskip
The positive roles played by chemotaxis in defense and repair  
become ambiguous in disease. Inflammation couples tightly to 
many disease processes \citep{hunter2012}, including in cancers, 
neurological diseases (e.g. multiple sclerosis, MS, and Alzheimer's disease, AD), 
heart diseases (e.g. artherosclerosis), diabetes, fibrosis {\em etc}.
Inevitably, the question emerges as to whether ``healthy'' chemotactic processes 
become corrupted. Cancer is, quite obviously, of enormous interest and chemotaxis 
has been associated with numerous events, including angiogenic growth, 
invasion/metastasis of cancer cells and cancer-immune system interactions: see 
\cite{roussos2011} for a review. Of course, unravelling 
the complex interactions and ascertaining whether a given chemotaxis event 
acts to delay or accelerate disease progress is far from trivial.

\subsection{Modelling}

Immune system dynamics have inspired volumes of models, with a 
number explicitly incorporating chemotaxis. Often, chemoattractant 
variables are discarded: for example, movement responses
to bacteria infections can be modelled by assuming cells orient 
with respect to bacteria density gradients rather than explicit 
chemoattractants. Earliest PKS-based formulations were made by Lauffenburger and
colleagues \citep{lauffenburger1979a,lauffenburger1979b,lauffenburger1983,lauffenburger1984}:
in a patterning context, the model of \cite{lauffenburger1983} (see (\ref{D1}))
demonstrated that a reduction in immune cell chemotaxis could lead to 
expansion and persistence of localised bacterial infections. Other PKS models 
for immune system dynamics include that in \cite{lee2017}, where patterning 
was investigated for immune cells responding to a foreign 
antigen, via chemokine-mediated chemotaxis (\ref{D2}). In the context of viral
infections, a spatial model of HIV dynamics was developed in \cite{stancevic2013}, 
where classic SIR dynamics were extended to incorporate 
chemotaxis. Specifically, uninfected ``target'' cells  were attracted to 
virus-infected cells (and exposed to virus particles), see (\ref{D3}), and the 
resulting model could generate spatial patterns representing infection ``hot-spots''.

\begin{figure}[t!]
\begin{center}
{\includegraphics[width=1.0\textwidth]{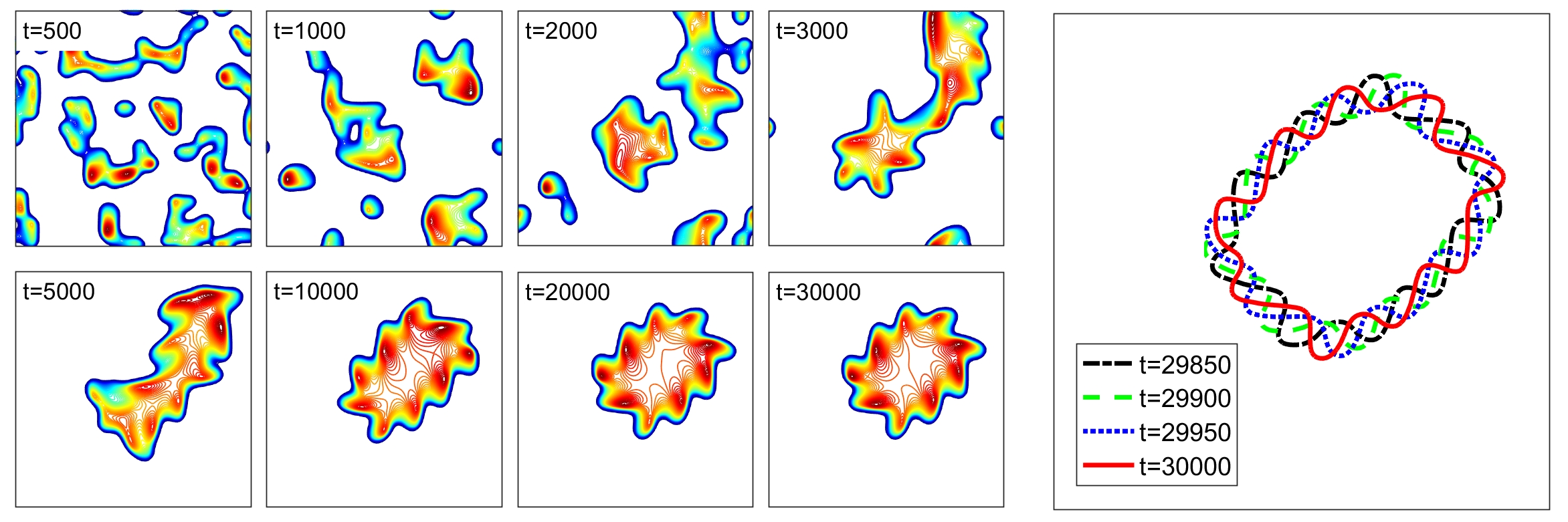}}
\end{center}
\caption{Evolution to a rotating star in the model (\ref{D4}), with parameters set at 
$D_m = 0.45, D_c = 1, k_1 = 0.5, k_2 = 0.6, q = 1, k_3 = 50$ and $\chi = 2.8$. Left: cell 
densities ($m$) plotted at the times shown, with contour lines from dark blue to dark red 
indicating increasing density. Right: illustration of the rotating structure, with contour
line for $m =100$ plotted at four successive time points.}\label{penner}
\end{figure}

\smallskip
Skin rashes provide portraits of inflammations, and capturing their form has 
been the motivation of modelling studies dating to the reaction-diffusion 
approach of \cite{segel1992}. The acute inflammation model proposed in \cite{penner2012} 
consisted of a generic immune cell population (e.g. macrophages) in the presence of a 
self-secreted chemokine and anti-inflammatory cytokine (\ref{D4}). A raft of peculiar 
patterns were observed, such as the ``rotating stars'' shown in Figure \ref{penner}. 
\cite{vig2014} specifically focussed on the rash {\em {Erythema migrans}},
an indicator of Lyme disease (transmitted via bacteria-infected tick bites). Their PKS model 
considered bacteria in stationary and motile forms and a 
population of chemotactic macrophages (\ref{D5}) and replicated various clinical rash 
morphologies, along with predicting the impact of antibiotic treatment.

\smallskip
Inflammation events in neurodegenerative disorders, including MS and AD,
have been investigated via a number of PKS models. The model in \cite{luca2003}
explored whether chemotactic microglia could be a driver in the formation of
senile placques, the lesions of proteins, degenerating neurons and glia cells that characterise 
AD. A combination of long range repulsion and short range attraction was shown to 
generate aggregates with the short wavelength clinical presentation. 
In Bal{\'{o}}’s concentric sclerosis\footnote{A rare form of MS, where the brain's oligodendrocytes 
are primary disease targets.}, a pathological concentric ring pattern is generated
as neurons lose their protective myelin sheaths (Figure \ref{figure1}f). 
A model featuring chemotactic macrophages, a pro-inflammatory chemoattractant and oligodendrocytes 
(see (\ref{D7})) was shown to replicate these rings and offered a potential origin 
\citep{khonsari2007}; extensions and a more detailed pattern 
analysis were considered in \cite{lombardo2017}. A highly detailed 
model (\ref{D8}) was developed in \cite{silchenko2015} to describe acute microglia 
responses, incorporating the interactions between microglia and attracting and/or 
repelling molecular substances (ATP, ADP, AMP and adenosine). Applied to the 
localised scarring generated at electrode implantation sites, the model generated
stable aggregates and anti-scarring strategies were considered. 

\smallskip
Angiogenesis has offered considerable modelling scope and a paragraph
is insufficient to cover all contributions: several reviews cover these models in
both tumour and wound induced angiogenesis \citep{mantzaris2004,scianna2013,flegg2015,heck2015}. 
Regarding pattern formation, the PKS model in \cite{orme1996a} was of 
classic form (\ref{D9}), featuring haptotactic-guided endothelial cells that respond to 
secreted fibronectin and initiate new sprouts along existing capillaries. Later
models, such as \cite{levine2001a}, greatly expanded the detail of cellular/molecular control.
Other angiogenesis models have focussed on the wavelike extension of 
vessels in response to tumour angiogenic factors (TAFs). Of these, the 
``snail-trail'' approach in \cite{balding1985} (inspired by models for fungal 
networks, \citealt{edelstein1982}) has proven popular: capillaries are 
compartmentalised into separate tip/vessel variables (\ref{D11}), with 
tip cells responding to TAF gradients and laying vessels as they 
migrate. Other early PKS-type models that focus on the extension of endothelial cells
include that of \cite{chaplain1993}, featuring an 
attractant that diffuses (and degrades) from a boundary-source and an endothelial 
population that migrates and proliferates in response.

\smallskip
While they offer analytically tractable caricatures of the angiogenesis
process, capturing the fine-scale structure of microvessels is difficult
within coarsescale/continuous PKS models. Dating to fundamental 
studies of the 1990s (e.g. \citealt{stokes1991,anderson1998}), a large number of
detailed agent-based/hybrid methods have been formulated, 
with their capacity to generate experimentally measurable outputs leading to 
a growing symbiosis between theory and experiment (see \citealt{bentley2013} for 
a review). The hybrid approach of \cite{anderson1998} is 
highlighted for its direct connection with PKS systems. Specifically, a 
continuous PKS-type system was proposed to describe tip cell densities (and other factors) 
and subsequently discretised in space to stipulate movement rules for an 
equivalent stochastic cellular automaton model. The capacity to generate realistic-looking 
network structure, as well as extend to include factors such as blood flow, has 
facilitated expansion of this approach and integration with experimental studies \citep{machado2011,mcdougall2012}. Fully continuous frameworks therefore 
remain a convenient method for formulating base models and ongoing work 
substantiates their link to underlying discrete processes, 
for example see \cite{pillay2017} in the context of snail-trail concepts.

\smallskip
PKS models have been proposed in other aspects of tumour growth and, again, 
detailing all models would demand a separate review: we focus on those where 
spatial patterning emerges. The model in \cite{owen1997} investigated
tumour-immune interplay, with macrophage chemotaxis in response to tumour-secreted 
factors (\ref{D12}). A complex pattern of invasion was observed, including irregular 
spatio-temporal ``tumour-clumps'' that could generate heterogeneous 
invasion forms; the extension in 
\cite{donofrio2012} incorporated chemorepulsion in response to 
immune system effectors. Tumour-macrophage dynamics have also been the
focus of \cite{knutsdottir2014}, where interactions between tumour and 
macrophage populations (\ref{D13}) were explicitly accounted for via 
a combination of ``autotaxis'' (taxis to a self-secreted chemical) and 
``parataxis'' (taxis to a chemical secreted by a separate cell population) 
terms. \cite{chaplain2006}
developed a sequence of tumour invasion models (\ref{D14}) that incorporated essential
tumour/matrix interactions, with chemotaxis of tumour cells to uPA protease
shown to induce complex/chaotic spatio-temporal clumping phenomena.

\section{Applications in ecology}

\subsection{Background}

Organisms frequently move in response to chemicals, self-evident from our 
reactions to enticing or noxious odours. Tracking dogs domesticated by our
prehistoric ancestors reveal a long appreciation for animal senses \citep{morey1994} and 
Aristotle in the {\em History of Animals}\footnote{An important translation 
was given by D'Arcy Wentworth Thompson, both pioneering mathematical biologist and 
highly-regarded classicist \citep{aristotlethompson}.} recounts various stories, such as the use of 
repellents to disperse ants.
Numerous orienteering paradigms have long been surmised to rely (partly) on 
odour sensing, including ant trail formation \citep{bonnet1779,rennie1831},
salmon homing \citep{trevanius1822} and moth mate location. Inspired by 
numerous male moths arriving at a room containing a captive female, 19th century naturalist Jean-Henri 
\cite{fabre1879,fabre1921} hypothesised her secretion of a powerful aroma. 
In one crucial experiment\footnote{In the course of his experiments, Fabre learned why 
you should not place a rare specimen in a tank containing a preying mantis. The
moth bore the brunt of the error.}, males ignored the visible female transferred to 
a new cage, attracted instead to the lingering smell of her previous perch. It is now 
firmly established that moths and many other species emit powerful pheromones that enable 
long-range communication.
 
\smallskip
Taxis responses are probable for many organisms: bilateral
olfactory organs (antennae, forked tongues, nostrils) 
suggests instantaneous spatial gradients could, in principal, be detected.
To varying certitude, chemotaxis has been proposed in soil nematodes 
\citep{lockery2011,rasmann2012}, the fruit 
fly {\em {Drosophila melanagaster}} in both larval \citep{gomez2012} and adult \citep{gaudry2012} stages, bees \citep{martin1965}, snakes and other reptiles \citep{schwenk1994}, moths 
and butterflies \citep{farkas1972}, various fish \citep{daghfous2012} including catfish \citep{johnsen1980} and sharks \citep{mathewson1972}, 
and lobsters \citep{reeder1980}. Lack of certainty stems from the difficulty of 
precise tests: environments are complex, plumes subject to turbulence
and contributions from other sensory inputs must be eliminated.

\smallskip
Our best understanding emerges in
small organisms. Nematodes, such as {\em C. elegans}, have received particular interest and various chemoattractants have been identified \citep{ward1973}. {\em C. elegans} climbs an attractant gradient 
via straight(ish) runs (generated by body undulations) alternating with ``pirouettes'' (reorientations). Analogous to {\em E. coli}, a similar klinokinesis behaviour 
has been identified in which pirouette frequency varies with the rate of concentration 
change \citep{pierce1999}. Yet this is further augmented by klinotaxis: during undulations, the 
head swings side to side and small directional shifts occur according to the attractant 
gradient \citep{iino2009}. Similar klinokinesis/klinotaxis combinations 
are suggested for larvae of the fruit fly {\em Drosophila melanogaster} \citep{gomez2012}.
An ability to orient directly to an instantaneous spatial gradient (tropotaxis) has 
been identified in adult {\em Drosophila}\footnote{Chemotaxis in fruit flies was 
investigated much earlier
\citep{severin1914}, exploring Kerosene as a trap for the Mediterranean Fruit 
Fly {\em Ceratitis capatata}. However, this species derives from 
a completely different family of fruit flies.} during both walking \citep{borst1982} and flying 
\citep{duistermars2009}: concentration differences are measured via the separated olfactory 
antennae. The extent to which this contributes to normal behaviour remains
debatable, as other odour-following strategies exist such as 
flying upwind (anemotaxis) when an odour is sensed \citep{gomez2012}. For mammals there
are conflicting findings, although a body of work suggests rats 
\citep{rajan2006}, and possibly even humans \citep{von1964,porter2007}, can `smell in stereo':
localising odours in a single sniff through the bilateral input provided by 
separated nostrils.

\smallskip
Self-organisation is widespread in ecology, with numerous species known 
to swarm, school, flock, herd {\em etc}. Chemically-mediated examples
lead to social insects, such as ants and their capacity to generate intricate trail 
networks \citep{czaczkes2015}, see Figure \ref{figure1} (g). Briefly rubbing a finger 
across a trail can bring ants to a standstill, and 18th century zoologist Charles 
\cite{bonnet1779} surmised that ants used odours to mark the route. The modern view 
developed from findings of the 1960s: \cite{wilson1962} showed that the different 
pheromone trails at a nest would recruit ants according to their concentration, while  \cite{hangartner1967,hangartner1969} showed 
that branch selection was proportional to its pheromone concentration 
and that ants release more pheromone if the food source is of higher quality. This suggested 
a powerful feedback process in which trails leading to the best sites are most strongly 
signalled, allowing the colony to efficiently harvest its surroundings (for a 
general review, see \citealt{czaczkes2015}). Further examples of chemically-mediated group 
aggregation can be found in species such as tent caterpillars \citep{fitzgerald1999}.

\subsection{Modelling}

\begin{figure}[t!]
\begin{center}
{\includegraphics[width=1.0\textwidth]{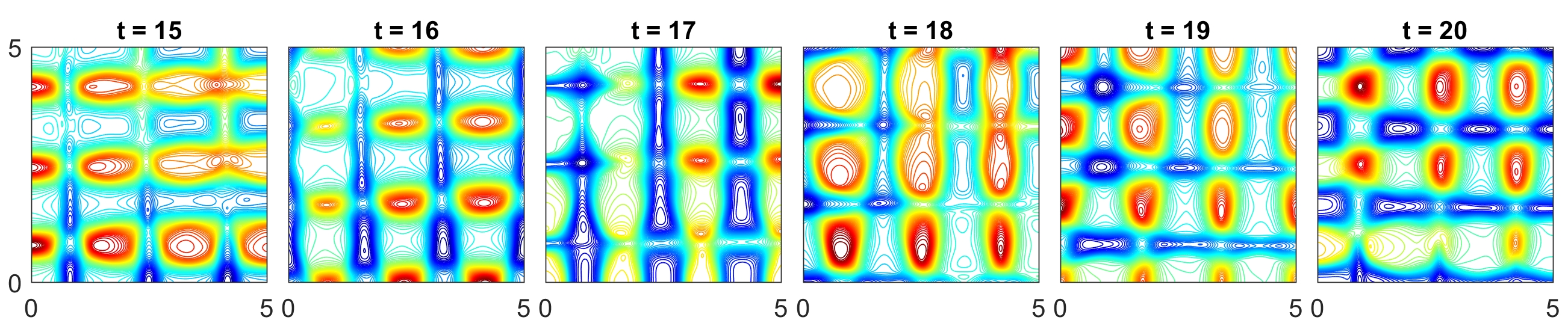}}
\end{center}
\caption{Oscillating patterns in the forager-scrounger model (\ref{E2}). Frames show forager density,
with contour lines blue to red reflecting increasing density. Parameters set at $D_f = D_s = 1$, 
$D_r = 0.1$, $\chi_f = \chi_s = 10$, $G = H = 0$ and $k_1 = 8.05, k_2 = 8, k_3 = 0.05$. The 
population is equally composed between foragers and scroungers.}\label{foragersims}
\end{figure}

PKS terms incorporated within ecological models are often 
phenomenologically motivated,
for example ``preytaxis'' fluxes to model predator 
movement responses to a prey distribution \citep{kareiva1987}.
Incorporating preytaxis into biological invasion models suggests it
can potentially accelerate invasion, by drawing predators away from the 
leading edge and giving outlying prey a greater survival chance \citep{lee2008}. In the
context of pattern formation prey-taxis acts to stabilise, shifting the system
towards spatial uniformity rather than group-forming behaviour \citep{lee2009}: 
intuitively, predators diminish their attractor (the prey) rather than 
reinforcing it. An intriguing extension, though, has been suggested in 
\cite{tania2012} where a PKS model (\ref{E2}) was used to describe ``forager/scrounger'' 
systems. Here, the forager (predator) searches for food (prey), modelled via
food gradient following. The scrounger instead exploits the efforts of the 
forager, following the forager gradient rather than actively seeking its own food:
interactions of this kind are observed, for example, in certain seabird populations. 
Surprisingly, the seemingly innocuous addition of scrounging can act as a destabiliser, 
breaking homogeneity to generate temporally-oscillating spatial patterns 
(Figure \ref{foragersims}).

\smallskip
PKS models have also been proposed for specific ecological systems. \cite{pearce2007} 
developed a multi-species host-parasitoid system (\ref{E3}), 
including two parasitoid wasps, their hosts (caterpillars of the large and small 
cabbage white butterflies) and a chemical secreted by feeding hosts. 
Wasps attracted by this factor accumulate at feeding sites, 
with sufficiently powerful responses inducing spatially structured, potentially chaotic, 
distributions. PKS models have 
also been applied to describe ant trail formation \citep{ramakrishnan2014,amorim2015}, with 
chemotaxis used to describe pheromone trail following by foraging ants (\ref{E4},\ref{E5}).
Pheromone-mediated movement has been added into various models of mountain pine beetle 
infestation, a destructive species that exploits (and kills) trees during its reproduction. 
Successful attacks demand pine beetle accumulation, coordinated via their secretion of powerful
aggregation-inducing pheromones. A detailed PKS-type model (\ref{E6}) described airborne 
and nesting beetles, their secreted pheromones, the density of tree attack holes for 
colonising beetles, chemicals 
(kairomones) released by host trees and the tree's resin capacity/outflow rate, see
\cite{powell1996}. While the dynamics of initial outbreaks were
shown to be determined by the distribution of host trees, later dynamics were
governed by the aggregation-inducing process of pheromone secretion/attraction \citep{logan1998}.
A reduced version of the model (\ref{E7}), featuring nesting and flying beetles and the aggregation 
pheromone, has been analysed in \cite{powell1998} and \cite{strohm2013}.

\section{Applications in the social sciences}

Core applications of PKS models obviously rest in biology, but we conclude the 
review component with a brief digression into the social sciences. Taxis
responses here are highly phenomenological, capturing some broad behaviour
rather than a specific movement response to a chemical gradient. For example, a PKS 
model developed in \cite{neto2015} was framed within an economic setting, addressing 
the interaction between ``capital'' and ``labour'' in economies: in essence, greater 
labour generates more capital and capital attracts labour, the latter explicitly 
modelled via a PKS-taxis flux of labour towards capital. The resulting 
model, see (\ref{F1}), was used to show how labour and capital co-localise 
and form dominating economies, as well as the potential for complex and 
unpredictable temporal behaviour to arise.

\begin{figure}[th!]
\begin{center}
{\includegraphics[width=0.9\textwidth]{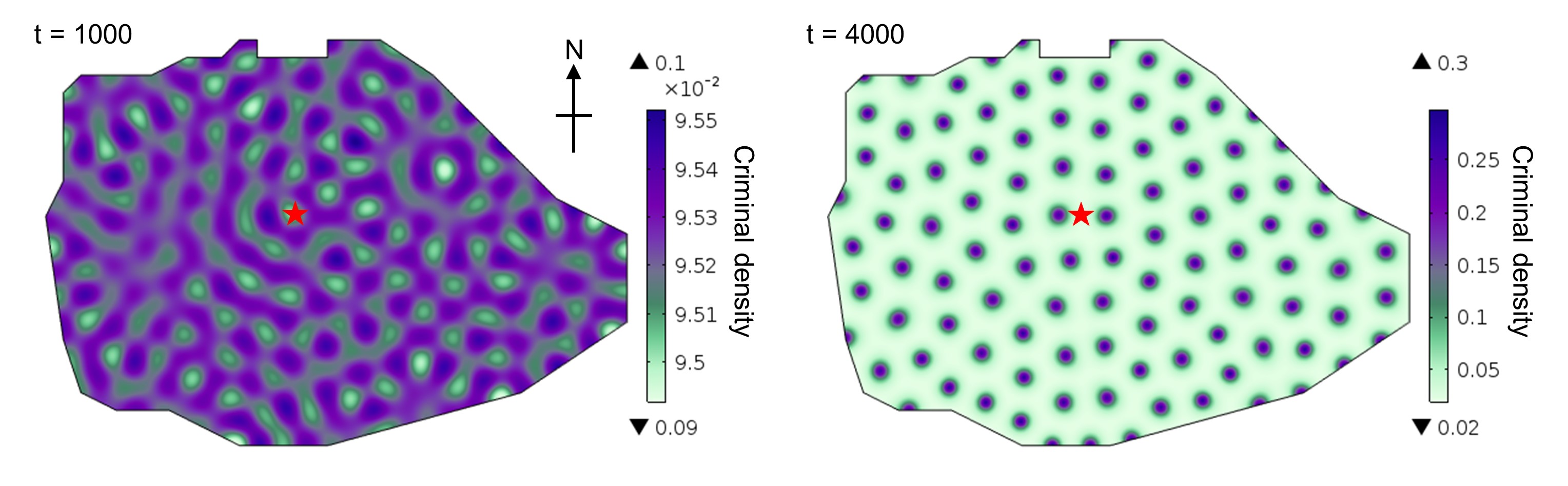}}
\end{center}
\caption{Crime hotspot model (\ref{F2}), where criminal density is plotted at $t=1000$ and 
$t=4000$ (green to purple reflects increasing criminal density) and parameters are set at 
$k_1 = 0.05, k_2 = 0.5, k_3 = 0.05, D_c = 100, D_a = 2$ and $b = 0.025$. Here, equations
have been solved on a spatial region approximately describing the outline of 
Edinburgh (red star reflects the position of Edinburgh Castle).} \label{figcrime}
\end{figure}

\smallskip
Crime modelling has received significant interest and statistical analyses of 
criminal data reveal spatio-temporal structuring such as transient or long-lasting 
``hotspots'' of elevated activity.  \cite{short2008,short2010b} sought to 
explore how these hotspots arise. An 
agent-based model was proposed for residential crime 
activity, based on  ``repeat-victimisation'' and ``broken-windows'' phenomena: 
burgled properties and their neighbours are statistically likely to suffer repeat attack in a 
period following a crime. Mobile criminal agents responded to an ``attractiveness'' field, 
representing the desirability of properties. In a continuous limit a 
PKS-type model (\ref{F2}) was obtained for criminal density and the attractiveness field, with 
biased movement towards attractive locations generating the 
taxis term. Crimes were committed at a rate according to local attractiveness,
modifying this field both locally (repeat-victimisation) and nearby 
(broken-windows) in the process. The resulting model had the essential structure of an autotaxis
system, so that spatial pattern (corresponding to hotspots) could occur 
even under initially homogeneous attractiveness (see Figure \ref{figcrime}): hotspots can 
emerge in cities without 
historically ``bad neighbourhoods''. While the concept of attractiveness is 
obviously difficult to quantify, the model successfully shows how the empirical 
observations can lead to complex spatial structuring and provides a means for 
exploring effective deployment of anti-crime measures. Various extensions have been considered \citep{short2010a,jones2010,short2010b,pitcher2010,zipkin2014,gu2017}, 
in particular to include policing (as one example, see model (\ref{F3})) and we refer 
to \cite{dorsogna2015} for a review of this developing field. 

\section{Academic clique formation}

The above sections reveal a widening usage of PKS equations, and we 
conclude with a light-hearted and novel case study. Specifically, we 
formulate a model for {\em research drift}, defined as the 
tendency of academics to alter their line of research over time. 
We show that biased drift according to perceived problems 
can generate an {\em academic clique} (a cluster of researchers working on 
the same topic) and ``hot'' research topics. The number of mathematicians currently 
analysing chemotaxis models would appear to be an apt example.

\subsection{Model equations}

We develop a model for the rate of change of academics, $a(\bx,t)$,
at time $t$. Here, $\bx \in \Omega$ is not position in standard 
three-dimensional space, but a specific {\em line of research}
in some {\em research field} $\Omega$. We assume $\bx$ is a continuous 
real variable, i.e. $\bx \in \Omega \subset \RR^n$: if mathematical biology 
was the research field, modelling tumour invasion or angiogenesis 
could be regarded as two separate, but close, points in $\Omega$\footnote{At its most 
general $\Omega$ would cover the full spectrum of academic disciplines (maths, 
physics, biology, humanities, social sciences {\em etc}) and is therefore 
of high dimensionality. This validates any theoretical attempts  
to prove results for dimensions $>3$...}. We assume academics can do two 
things\footnote{Insert a joke here...}:
\begin{enumerate}
\item[(S1)] solve problems;
\item[(S2)] change their line of research over time (research drift);
\end{enumerate}
Supposition (S1) implies that the number of problems is an important variable: if 
all problems are solved there is no point doing research. Of course, problems are 
of varying difficulty\footnote{Showing
$a^n + b^n = c^n$ ($a,b,c,n\in \ZZ$) has no solutions for $n>2$ was considerably harder 
than showing it does for $n = 2$...} and, to account for this variation, we compartmentalise
problems into harder ($h(\bx,t)$) and easier ($e(\bx,t)$) problems. (S2) implies a movement 
flux in which academics move through their research field. Following classical 
conservation arguments, 
\[
\pdifft a = - \nabla \cdot {\bf {J}}_a + f(\cdot)\,,
\]
where flux ${\bf {J}}_a(\bx,t)$ describes research drift and $f$ is the population 
change. We assume any movement is local -- a nonlocal (e.g. integral) 
formulation could allow scholars to radically change research, 
but we discount this at present. Research drift is assumed to derive from: 
(i) a random component; (ii) directed components towards areas with 
many identifiable problems\footnote{``Problemataxis''}. The former describes 
``chance encounters'' in a nearby topic (e.g. inspired by a chance-found paper), while 
the latter accounts for actively seeking areas with questions to solve (e.g. 
noting hot topics at conferences). The function $f$ is likely to be non-zero, 
since researchers can enter (PhD students...) or exit (retiring or moving into university 
management...). Generally, $f$ would depend on both the number of academics and 
problems: problems to solve implies papers to write implies  
grant success implies new students; no more problems, and academics may exit research. 
Taken together, we assume the following PKS equation for $a(\bx,t)$:
\begin{equation} \label{academics}
\pdifft{a} = \nabla \cdot \left[ D \nabla a - \chi_{e} a \nabla e - \chi_{h} a \nabla h
\right] + f(a,h,e)\,.
\end{equation}
$D, \chi_{e}$ and $\chi_h$ are likely to be functions of both 
the number of academics and number of problems, however for simplicity we 
shall set them to be constants. $\chi_{e,h}$ are measures of
attraction towards problems and how rapidly a research line can be altered.

\smallskip
Problem variables are governed by the following factors:
\begin{enumerate}
\item[(S3)] problems are solved at a rate that increases with the number of academics;
\item[(S4)] solving one problem can create further problems;
\item[(S5)] problems ``transfer'' to nearby areas. 
\end{enumerate}
(S3) follows a ``more hands make light work'' presumption: problems will be 
solved more quickly if more academics work in the area. (S4) presumes that 
solving one problem can lead to new ones, e.g. questions raised in the
discussion of a paper/conclusions of a talk. (S5) accounts for how questions 
raised in one field can generate a similar question in a nearby area. 
Here, we set
\begin{eqnarray}
\pdifft e & = & D_e \nabla^2 e - g_e(a) e + \gamma_{eh} g_h(a)h+ \gamma_{ee} g_e(a)e -\delta_e e\,, \label{easy}\\
\pdifft h & = & D_h \nabla^2 h - g_h(a) h + \gamma_{hh} g_h(a)h+ \gamma_{he} g_e(a)e -\delta_h h\,. \label{hard}
\end{eqnarray}
We note that ``transfer'' has been modelled simplistically here, via diffusion terms, 
although (S5) more accurately suggests problem creation in adjacent areas. Problems 
are solved at rates $g_{i} (a), i \in \left\{e,h\right\}$, and solving a 
problem of type $j$ is assumed to create $\gamma_{ij}$ new problems of type $i$. The 
linear decay terms account for raised problems being ``forgotten'': e.g. nobody reads 
the discussion :-(

\subsection{Functions and parameter constraints}

We solve (\ref{academics}-\ref{hard}) under choices
$f=0$ and $g_{i} = \frac{\alpha_{i} a}{1+\kappa_{i} a}$ for $i\in\left\{e,h\right\}$. 
The simplification $f = 0$ presumes a population of {\em selfish immortal academics}: they 
never retire, never 
move into administration and never supervise new students\footnote{Many academics may 
want this. More charitably, $f=0$ implies 
each exiting academic is replaced by one new researcher.}. The choice of $g$'s assumes 
problems are solved at a rate that increases, but eventually saturates, with the number of 
academics: a ``law of diminishing returns'', where increasing numbers leads to 
overlapping work and proportionally lower rate gain of problem solving.

\smallskip
Parameter restrictions are motivated by the following set of considerations.
\begin{enumerate}
\item[(P1)] $\chi_e > \chi_h$. Academics are more attracted to areas with 
easier problems. This reflects two factors: (i) ``publish or perish'', 
where quickly solving problems (and hence publishing) is considered the 
best method for rapid career progression; (ii) harder problems are 
difficult to identify. We specifically set $\chi_h = 0$ here.
\item[(P2)] $\alpha_e \gg \alpha_h > 0$. The maximum rate of solving easier problems
is significantly faster than solving hard problems, although the latter is nonzero.
\item[(P3)] $\delta_e \gg \delta_h$. Easier problems are far more likely to 
be forgotten than harder problems: `Fermat's last theorem was remembered 
more than three centuries later, but no-one remember the problem I posed during 
the talk at XXXX last year'. We specifically set $\delta_h = 0$ here.
\item[(P4)] $\gamma_{eh} \gg \gamma_{hh}, \gamma_{ee}, \gamma_{he}$. In other words, 
solving a hard problem can generate lots of easy problems: devising a new 
analytical/experimental technique that can be adopted for similar problems. We 
specifically consider $\gamma_{eh}>0$ and $\gamma_{hh} = \gamma_{ee} = \gamma_{he} = 0$ here.
\end{enumerate}
Initially we suppose academics are uniformly distributed across the field, 
$a(\bx,0) = a_0$, so that there is no initial clustering. We further 
start with hard problems only: easy problems only emerge when some hard 
problems are solved. Thus, we take $e(\bx,0)=0$ and set 
$h(\bx,0) = h_0(1+r(\bx,0))$ where $r(\bx,t)\in[-0.01,0.01]$ represents 
a small random variable. At boundaries we impose zero-flux 
conditions for both the academics and problems: i.e. neither can exit 
their research field\footnote{Polymaths of the past, such as D'Arcy Wentworth Thompson, 
are taken to be rare.}.

\subsection{Dynamics and clustering}

\begin{figure}[t!]
\begin{center}
{\includegraphics[width=\textwidth]{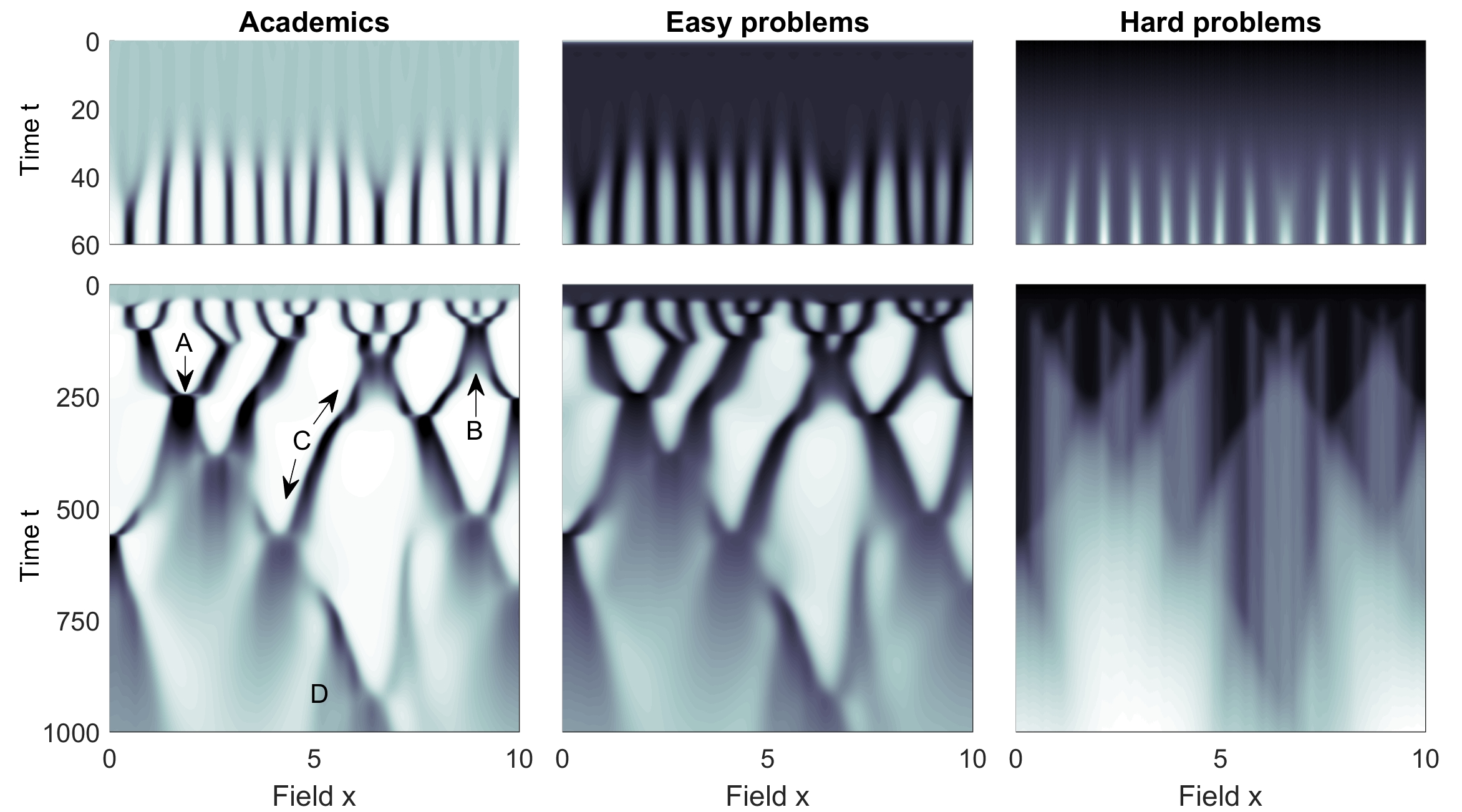}}
\end{center}
\caption{Simulations of the research drift model (equations (\ref{academics}-\ref{hard})), with 
functions and parameter constraints as described in the text. Grayscale from white to 
black represents increasing density of academics or problems. Top row: incipient stages of 
pattern formation, where an initially uniform distribution of academics self-organise into 
distinct ``cliques'' as progress is made on solving problems. Bottom row: spatio-temporal 
dynamics over longer timescales, with dynamics showing: (A) convergence; (B) divergence; 
(C) clique drift; and (D) eventual field death. In these simulations we set $D_A = D_E = 0.01, D_H = 0$, 
$\chi_E = 0.1, \chi_H = 0$, $\alpha_e = 1$, $\alpha_h = 0.001$, $\kappa_h = 0.1$, 
$\kappa_e = 0.01$, $\gamma_{eh} = 1, \gamma_{hh} = \gamma_{ee} = \gamma_{he} = 0$.}\label{clustering}
\end{figure}

We solve (\ref{academics}-\ref{hard}) subject to the functions, parameter 
constraints, initial and boundary conditions above. Note that for simplicity we 
restrict the academic field to a one dimensional line $[0,L]$; points
$0$ and $L$ can be interpreted as the most distantly related topics in a
research field. For low values of $\chi_e$ (academics do not tend to
alter their research according to problems), no patterning emerges and 
the distribution of scholars remains uniform. For larger values, however,
a self-organisation process occurs with typical simulations shown in Figure 
\ref{clustering} for (a) shorter, and (b) longer timescales. Briefly, the underlying 
randomness leads to a variable rate of progress at solving hard problems. Solving harder problems, 
however, then creates numerous easier problems and fuels a ``gold-rush'' of 
eager academics from nearby areas. This in turn quickens the rate at which harder problems 
are solved, reinforcing the process. 
Thus, we have the essential feedback for 
taxis-based self-organisation.

\smallskip
Over longer timescales we observe more complicated phenomena as follows.
\begin{trivlist}
\item (A) Research convergence, where closely related groups ``join forces'' and start 
working on a common topic.
\item (B) Research divergence, where a group splits into separate research lines.
\item (C) Group drifting, where a group of academics steadily shifts its
line of research as it searches for new problems to work on.
\item (D) Field death. Eventually, all problems are solved and clusters disperse. 
\end{trivlist}

\section{Perspectives}

More than a century has passed since the earliest identification of chemotaxis, and more than half 
a century since their first mathematical models. In the following years a vast 
literature has developed, and chemotactic behaviour has been incorporated into numerous models to address 
numerous questions in numerous fields. Moreover, a large mathematical literature has emerged,
leading to an elegant understanding of their mathematical properties. As of late 2017, the 
first paper of \cite{keller1970} has attracted more than 2,000 citations 
(Google Scholar). 

\smallskip
At what point does a model stop being useful? Paraphrasing Mae West\footnote{``I'm no 
model lady, a model's just an imitation of the real thing''.}, a model's just an imitation 
of real life so the obvious answer is when it fails to imitate. Inevitably, this depends 
on the problem being addressed, but, in all discussed areas, recent applications 
have been made of the PKS framework. This certainly suggests their ongoing relevance
but it is noted that for the vast majority of studies, model {\em fitting} has generally been 
performed at a qualitative or semi-quantitative level. As data becomes
increasingly nuanced, more stringent tests must be made via carefully constructed quantitative 
comparisons: in this regard the recent study of \cite{ferguson2016} provides a promising start.

\smallskip
Molecular/individual-scale understanding has advanced considerably and, with parallel 
advances in computational power, modelling has evolved via agent-based approaches 
and their ilk. These have proven enormously appealing, admitting intricate and
tailored detail that connects easily to biological data: unsurprisingly, all 
applications considered have witnessed significant modelling of this nature. 
Nevertheless, these approaches rely heavily on computation while 
continuous models can admit deeper analysis and broader-scale 
enlightenment. Keeping pace with these individual models, however, will demand
that continuous models evolve to import key microscopic elements.
For example, phenomena such as ``collective chemotaxis'' \citep{theveneau2010},
where clusters performs chemotaxis more efficiently 
than isolated cells, appears to fundamentally depend on microscale interactions
between cells: can appropriate PKS-type models be derived to describe these phenomena?
Other areas for fundamental modelling include accounting
for cell polarization or heterogeneous 
populations, where two or more populations exist with distinct movement 
properties. In terms of the latter, one approach (taken by several models in the compendium) 
is to simply include multiple populations with standard fluxes. This may be 
reasonable at relatively low densities, but in tightly packed populations (e.g. 
spore/stalk cells in the slug), movement of one population could clearly 
influence the other. A number of attempts have been made to model such 
behaviour, e.g. \cite{painter2003,painter2009,simpson2009,johnston2015}.

\smallskip
The number of recent applications, particularly in areas such as 
ecology and the social sciences, suggests that the well of problems 
is far from dry. Indeed, its
limits may rest only in our imagination and, to illustrate, we have 
considered a novel application in the social sciences. Taxis here is a hazier concept,
and the broad nature of PKS-type fluxes is
an attractive way of characterising the essence of a phenomena. While our model has 
been constructed in a somewhat playful manner, there is obvious interest in the 
positives and negative implications of social group formation: not just academically, but 
school or office cliques, political opinion, groupthink, religious circles, 
social networking sites {\em etc.} \citep{backstrom2006}. For example,
the ideas here could be adapted to explore the drivers of extremism, 
for a population structured according to opinion and fluxes driving 
shifting political thought.

\noindent
Our modelling provides an intuitive explanation for how academic 
cliques form: progress on harder problems leads to
easier problems, attracting scholars. Progress quickens and so forth. Over time, 
various phenomena emerge including the convergence and divergence of groups 
or group drifting across the research field. Our principal aim has been 
demonstrative, but various factors could warrant greater analysis: including nonzero 
kinetics to account for incoming PhD students {\em etc}; population heterogeneity 
(e.g. ``lone wolf'' academics that avoid crowded areas); including the impact from
measurable factors such as research funding and publications {\em etc}. 

\noindent
In our model, the number of problems, eventually, runs dry. Research clustering 
disappears and the field dies. In a population of selfish immortal academics, 
they disperse and are condemned to an eternity of writing review articles.

\medskip
{\bf Acknowledgements.} KJP would like to thank the Politecnico di 
Torino for a Visiting Professorship position, Thomas Hillen for highly 
constructive comments and Philip Maini for originally introducing him to the 
subject.

\appendix
\section{A compendium of PKS models for pattern formation} \label{compendium}

Note that in the following we generally use $D_i$'s to denote diffusion 
coefficients, $\chi_i$'s to denote chemotactic sensitivity coefficients 
and $k_i$'s as constants. We refer to the original articles 
for full details and model motivations.\\

\subsection{Models for {\em Dd} aggregation}

\cite{keller1970} model for {\em Dd} aggregation, for 
densities of amoebae ($a$), concentrations of acrasin ($c$), acrasinase ($p$) and complex ($q$): 
\begin{equation}\tag{A1}\label{A1}
\begin{array}{rcl}
a_t & = & \nabla \cdot \left[ D_a \nabla a - a \chi(a,c) \nabla c \right]\,, \\
c_t & = & D_c \nabla^2 c + a F(c) + k_{2} q - k_1 c p\,, \\
q_t & = & D_q \nabla^2 q + k_1 c p - (k_{2} + k_3) q\,, \\
p_t & = & D_p \nabla^2 p + (k_{2} + k_3 ) q  + a G(c,p) - k_1 p c \,. 
\end{array}
\end{equation}
Reduced (two variable) version:
\begin{equation}\tag{A2} \label{A2}
\begin{array}{rcl}
a_t & = & \nabla \cdot \left[D_a \nabla a - a \chi(a,c) \nabla c \right]\,, \\
c_t & = & D_c \nabla^2 c + a F(c) - H (c) c\,.
\end{array}
\end{equation}
\cite{hofer1995a} model for {\em Dd} aggregation, for cells ($n$), cAMP concentration ($u$) and fraction of active cAMP receptors ($v$):
\begin{equation}\tag{A3}\label{A3}
\begin{array}{rcl}
n_t & = & \nabla \cdot \left[ D_n \nabla n - \chi_0 \frac{n v^m}{k_3^m+v^m} \nabla u \right]\,,\\
u_t & = & D_u \nabla^2 u + k_1 \left( G(n) F(u,v) - (G(n)+k_2) k_6 u \right)\,,  \\
v_t & = & -k_9 u v + k_{10} (1-v)\,,
\end{array}
\end{equation}
where $F(u,v) = (k_4 v + v^2)(k_5+u^2)/(1+u^2)$, $G(n) = n/(1-k_7 n/(k_8+n))$. 

\subsection{Models for bacteria pattern formation}

\cite{woodward1995} model for salmonella pattern formation, for bacteria density ($u$) and aspartate ($v$):
\begin{equation}\tag{B1}\label{B1}
\begin{array}{rcl}
u_t & = & \nabla \cdot \left[ D_u \nabla u -  \frac{\chi u}{(1 + k_1 v)^2} \nabla v \right] + k_2 u \left( 1 - \frac{u}{s}\right)\,, \\
v_t & = & D_v \nabla^2 v + \frac{k_3 s u}{1 +  k_4 u} - \frac{k_5 v u^{p}}{1+k_6 v}\,.
\end{array}
\end{equation}
Note that $s$ represents a (constant) nutrient source.

\smallskip
\cite{tsimring1995} model for bacteria pattern formation, for motile bacteria ($m$), non-motile bacteria ($n$), nutrient ($f$), waste product ($w$), chemoattractant ($c$):
\begin{equation}\tag{B2}\label{B2}
\begin{array}{rcl}
m_t & = & \nabla \cdot \left[D_m \nabla m - \chi m \nabla c\right] + k_1 m^p \frac{f}{f+k_2} - m^3 - H_1(\cdot) m\,,\\
n_t & = & H_1(\cdot) m\,,\\
f_t & = & D_f \nabla^2 f - k_3 m f\,,\\
w_t & = & D_w \nabla^2 w + k_4 m f - k_5 w - k_6 m w\,, \\
c_t & = & D_c \nabla^2 c + H_2(\cdot) m - k_7 c\,.
\end{array}
\end{equation} 
Functions $H_1(\cdot)$ and $H_2(\cdot)$ are based on Heaviside functions,
respectively describing the switching to nonmotile form as starvation occurs 
and chemoattractant secretion according to the waste product. 

\smallskip
\cite{tyson1999a} model for bacteria pattern formation, for bacteria ($n$), nutrient (succinate, $s$) and chemoattractant (aspartate, $c$):
\begin{equation}\tag{B3} \label{B3}
\begin{array}{rcl}
n_t & = & \nabla \cdot \left[ D_n \nabla n - \frac{\chi n}{(k_1+c)^2} \nabla c \right]
+k_2 n \left(\frac{k_3 s^2}{k_4+s^2} - n\right)\,, \\
c_t & = & D_c \nabla^2 c + k_5 s \frac{n^2}{k_6+n^2} - k_7 n c\,, \\
s_t &= & D_s \nabla^2 s - k_8 n \frac{k_3 s^2}{k_4+s^2}\,.
\end{array}
\end{equation} 

\smallskip
\cite{polezhaev2006} model for bacteria pattern formation, for vegetative bacteria ($n$), anabiotic bacteria ($a$), chemoattractant ($c$) and nutrient ($s$):
\begin{equation}\tag{B4} \label{B4}
\begin{array}{rcl}
n_t & = & \nabla \cdot \left[ D_n \nabla n - n \frac{\chi}{(c+k_1)^2} \nabla c \right] + \frac{k_2 s }{s+k_3}n - k_4 n H(n)H(n+a-k_5)\,,\\
a_t & = & k_4 n H(n)H(n+a-k_5)\,, \\
c_t & = & D_c \nabla^2 c+ A(s) n - k_6 c\,, \\
s_t & = & D_s \nabla^2 s - \frac{k_2 s}{s+k_3}n \,,
\end{array}
\end{equation} 
where $A(s) = \left\{ \begin{array}{ll} k_7 s\,, & s \ge k_{8} \\ k_7 k_{8}\,, & s < k_{8} \end{array} \right.$ and $H(\cdot)$ represents the Heaviside function.

\smallskip
\cite{aotani2010} model for bacteria pattern formation, for active cells ($u$), inactive cells ($w$), chemoattractant ($c$) and nutrient ($n$):
\begin{equation}\tag{B5} \label{B5}
\begin{array}{rcl}
u_t & = & \nabla \cdot \left[ D_u \nabla u - u \nabla \frac{\chi_0 c^2}{c^2 + k_1^2} \right] + g(u) n u - \frac{k_4}{k_5+n} u\,,\\
c_t & = & D_c \nabla^2 c + k_6 u - k_7 c\,, \\
n_t & = & D_n \nabla^2 n - k_8 g(u) n u\,, \\
w_t & = & \frac{k_4}{k_5+n} u\,, 
\end{array}
\end{equation} 
for $g(u) = (1+\tanh(k_2 (u -k_3))/2$.

\smallskip
\cite{baronas2015} for {\em E. coli} pattern formation, for bacteria ($n$), nutrient (succinate, $s$) and chemoattractant (aspartate, $c$):
\begin{equation}\tag{B6} \label{B6}
\begin{array}{rcl}
n_t & = & \nabla \cdot \left[ D_n \nabla n - \chi n \nabla c \right]
+k_1 n \left(1 - n/k_2 s \right)\,, \\
c_t & = & D_c \nabla^2 c + \frac{k_3 n}{k_4+n} - k_5 c\,, \\
s_t &= & D_s \nabla^2 s - k_6 n \,.
\end{array}
\end{equation} 

\smallskip
Chemotaxis-fluid models \citep{hillesdon1995,tuval2005,chertock2012}) for {\em B. subtilis} pattern formation in thin fluid layers, for bacteria ($n$), oxygen ($o$) and fluid velocity vector field ($\bu$):
\begin{equation}\tag{B8} \label{B8}
\begin{array}{rcl}
n_t +\bu \cdot \nabla n & = & \nabla \cdot \left[ D_n \nabla n - \chi(o)  n \nabla o \right]\,, \\
o_t +\bu \cdot \nabla o & = & D_o \nabla^2 o - k_1 n F(o) \,, \\
\rho (\bu_t + \bu \cdot \nabla \bu ) &= & - \nabla p + \eta \nabla^2 \bu - n \nabla G \,,
\end{array}
\end{equation} 
and $\nabla\cdot \bu = 0$, with pressure $p$ and viscosity $\eta$. Gravitational force is modelled by 
$\nabla G = V_b g {\bf z} (\rho_b-\rho)$,  where ${\bf z}$ is the downward unit vector, $V_b$ is the volume of the bacteria and $\rho, \rho_b$ are respectively fluid and bacteria densities.

\subsection{Models for embryogenesis}

Chemotaxis model applied to embryonic pattern formation in chondrogenesis \citep{oster1989,maini1991,myerscough1998}, snake pigmentation \citep{murray1991} and feather morphogenesis \citep{lin2009}. Cells ($n$) and chemoattractant ($c$):
\begin{equation}\tag{C1}\label{C1}
\begin{array}{rcl}
n_t & = & \nabla \cdot \left[ D_n \nabla n - \chi n \nabla c \right] + k_1 n (k_2-n)\,,\\
c_t & = & D_c \nabla^2 c + \frac{k_3 n}{k_4 + n} - k_5 c\,.
\end{array}
\end{equation} 

\smallskip
\cite{michon2008} model for feather morphogenesis, for proliferating ($n$) and migrating ($m$) mesenchymal cells, activator chemoattractant ($u$) and inhibitor ($v$):
\begin{equation}\tag{C2}\label{C2}
\begin{array}{rcl}
n_t & = & \left\{ \begin{array}{cl} k_1 n (k_2 - n) & t \le t^* \\ -k_3 n & t > t^* \end{array} \right. \\
m_t & = & \nabla \cdot \left[ D_m \nabla m - \chi m \nabla u \right] + 
\left\{ \begin{array}{cl} 0 & t \le t^* \\ k_3 n & t > t^* \end{array} \right.\,,\\
u_t & = & D_u \nabla^2 u + \frac{k_4 m (1+k_5 u^2)}{(k_6+u^2)(1+v)} - k_7 u\,, \\
v_t & = & D_v \nabla^2 v + k_8 m u^2 - k_9 v\,.
\end{array}
\end{equation} 

\smallskip
\cite{painter2018} model for feather morphogenesis, for mesenchymal cells ($m$), activated epithelial ($e$), FGF ($f$) and BMP ($b$),
\begin{equation}\tag{C3}\label{C3}
\begin{array}{rcl}
m_t & = & \nabla \cdot \left[ D_m \nabla m - \chi m e^{-k_1 m} \nabla f \right]\,, \\
e_t & = & (k_2 W(x,y,t) G_1(m) + k_2 G_2(m)) (1-e) - \left(1-G_1(m)\right) \left( k_3 + k_4 b\right) e \,,\\
f_t  & = & D_f \nabla^2 f + k_5 e - k_6 f\,, \\
b_t & = & D_b \nabla^2 b + k_7 G_3(m) m - k_8 b\,,
\end{array}
\end{equation}
with priming wave $W(x,y,t)= k_9 (1+\tanh(k_{10} (t- y/k_{11})))$ and $G_i(m) = m^{p_i}/(\kappa_i^{p_i}+m^{p_i})$.

\subsection{Models in physiology and disease}

\cite{lauffenburger1983} model for tissue inflammation responses to infection, for a bacteria population ($b$) and 
phagocytising immune cells ($p$):
\begin{equation}\tag{D1}\label{D1}
\begin{array}{rcl}
p_t & = & \nabla \cdot \left [ D_p \nabla p - \chi p \nabla b \right] + k_1 + k_2 b - k_3 p\,, \\
b_t & = & D_b \nabla  b + \frac{k_4 b}{k_5+b} - \frac{k_6 b p}{k_7+b} \,.
\end{array}
\end{equation} 

\smallskip
\cite{lee2017} model for active immune cells ($m$), antigen ($a$) and chemokines ($c$):
\begin{equation}\tag{D2} \label{D2}
\begin{array}{rcl}
m_t & = & \nabla \cdot \left[ D_m \nabla m - \chi(c) m \nabla c \right] +k_1 -k_2 m a - k_3 m\,,\\
a_t & = & D_a \nabla^2 a + s(t,x) - k_3 m a - k_4 a\,, \\
c_t & = & D_c \nabla^2 c + k_5 m a - k_6 c\,. 
\end{array}
\end{equation} 
Note that $s(t,x)$ represents an antigen source (spatially localised).

\smallskip
\cite{stancevic2013} model for HIV infection spatial dynamics, for target cells ($n$), infected cells ($i$)  and virus ($v$):
\begin{equation}\tag{D3}\label{D3}
\begin{array}{rcl}
n_t & = & \nabla \cdot \left[ D_n \nabla n - \chi(n) n \nabla i \right] + k_1 - k_2 n v - k_3 n\,, \\
i_t & = & D_i \nabla^2 i + k_2 n v - k_4 i\,, \\
v_t & = & D_v \nabla^2 v + k_5 i - k_6 v\,. 
\end{array}
\end{equation}

\smallskip
\cite{penner2012} model for tissue inflammation, applied to skin rashes, for macrophages ($m$), a chemokine ($c$) and anti-cytokine ($a$):
\begin{equation}\tag{D4} \label{D4}
\begin{array}{rcl}
m_t & = & \nabla \cdot \left[ D_m \nabla m - \frac{\chi m}{(1+k_1 c)^2} \nabla c \right]\,, \\
c_t & = & D_c \nabla^2 c + \frac{m}{1+k_2 a^q} - c\,,\\
a_t & = & \frac{1}{k_3} \left(D_c \nabla^2 a + \frac{m}{1+k_2 a^q} - a \right)\,.
\end{array}
\end{equation} 

\smallskip
\cite{vig2014} model for {\em Erythema Migrans}, for macrophages ($m$), translocating bacteria ($b$) and stationary bacteria ($s$):
\begin{equation}\tag{D5}\label{D5}
\begin{array}{rcl}
m_t & = & -\nabla \cdot \left[ \chi m \nabla (b+s) \right] + k_1 (b+s) - k_2 m \,,\\
b_t & = & D_b \nabla^2 b + k_3 b + k_4 s - k_5 b - k_6 m b\,, \\
s_t & = & k_3 s  - k_4 s + k_5 b - k_7 m s\,. 
\end{array}
\end{equation} 

\smallskip
\cite{luca2003} model for plaque formation in AD, for
microglia ($m$) and regulatory chemicals ($b, c$):
\begin{equation}\tag{D6}\label{D6}
\begin{array}{rcl}
m_t & = & \nabla \cdot \left[ D_m \nabla m - \chi_b m \nabla b + \chi_c m \nabla c \right]\,,\\
{b}_t & = & D_b \nabla^2 b + k_1 m - k_2 b\,, \\
{c}_t & = & D_c \nabla^2 c + k_3 m - k_4 c\,.
\end{array}
\end{equation} 

\smallskip
\cite{khonsari2007} and \cite{lombardo2017} model for Bal{\'{o}}’s concentric sclerosis, for
macrophages ($m$), oligodendrocytes ($d$) and chemoattractant ($c$):
\begin{equation}\tag{D7}\label{D7}
\begin{array}{rcl}
m_t & = & \nabla \cdot \left[ D_m \nabla m - \chi(m) \nabla c \right] + k_1 m (k_2 - m)\,,\\
k_3 c_t & = & D_c \nabla^2 c + k_4 d + k_5 m - k_6 c\,, \\
d_t & = & \frac{k_7 m}{k_2 + m} m \left(k_8 - d \right)\,.
\end{array}
\end{equation}
Note that $\chi(m) = \chi_0 m (k_2 - m)$ in \cite{khonsari2007} 
and $\chi(m) = \chi_0 \frac{m}{k_2 + m}$ in \cite{lombardo2017}; $k_3 = k_5 = 0$ in \cite{khonsari2007}.

\smallskip
\cite{silchenko2015} model for microglia aggregation, for
microglia ($n$), and concentrations of ATP ($a$), ADP ($b$), AMP ($c$) and adenosine ($d$) and IL-1 ($e$):
\begin{equation}\tag{D8}\label{D8}
\begin{array}{rcl}
n_t & = & \nabla \cdot \left[ D_n \nabla n - \frac{\chi_1 n}{(k_1+a)^2} \nabla a  - \frac{\chi_2 n}{(k_2+b)^2} \nabla b  - n \chi_3 F(d) \nabla d + n \chi_4 \nabla e \right]\,, \\
a_t & = & D_a \nabla^2 a + k_3 n + \frac{k_4 a}{a+k_5} - \frac{k_6 a}{a+k_7} - \frac{k_8 a}{a+k_9}\,,\\
b_t & = & D_b \nabla^2 b + \frac{k_6 a}{a+k_7} - \frac{k_{10} b}{b+k_{11}}\,,\\
c_t & = & D_c \nabla^2 c + \frac{k_{10} b}{b+k_{11}} + \frac{k_8 a}{a+k_9} - \frac{k_{12} c}{c+k_{13}(1+k_{14} b)}\,, \\
d_t & = & D_{d} \nabla^2 d + \frac{k_{12} c}{c+k_{13}(1+k_{14} b)} - \frac{k_{15} d}{d+k_{16}}\,, \\
e_t & = & D_{e} \nabla^2 e - k_{17} e + k_{18} n + k_{19}\,.
\end{array}
\end{equation}
Note that $F(d) = (k_{20} - k_{21} d^2)/(k_{22}^2+d^2)$.

\smallskip
\cite{orme1996a} model for capillary sprout initiation, for endothelial cells ($n$) and fibronectin ($c$):
\begin{equation}\tag{D9}\label{D9}
\begin{array}{rcl}
n_t & = & \nabla \cdot \left[ D_n \nabla n - \chi n \nabla c \right] + k_1 n (k_2-n)\,,\\
c_t & = & D_c \nabla^2 c + \frac{k_3 n}{k_4 + n} - k_5 c\,.
\end{array}
\end{equation}

\smallskip
\cite{orme1996a} model for capillary branching, for endothelial cells ($n$), matrix density ($r$) and adhesion sites ($a$):
\begin{equation}\tag{D10}\label{D10}
\begin{array}{rcl}
n_t & = & \nabla \cdot \left[ D_n \nabla n - \chi_1 n \nabla a + \chi_2 n \nabla r \right]\,, \\
a_t & = & \nabla \cdot \left[D_a \nabla a + \chi_2 a \nabla r\right]  + k_1 n - k_2 a\,, \\
r_t & = & D_r \nabla^2 r + k_3 n - k_4 r\,.
\end{array}
\end{equation}

\smallskip 
\cite{balding1985} snail-trail model for angiogenic sprouting, for tip cells ($n$) and vessel cells ($v$) and chemoattractant ($c$):
\begin{equation}\tag{D11}\label{D11}
\begin{array}{rcl}
n_t & = & - \nabla \cdot \left[ n \chi \nabla c \right] + k_1 c v - k_2 n v\,, \\
v_t & = & \left| n \chi \nabla c \right| -k_3 v\,, \\
c_t & = & D \nabla^2 c.
\end{array}
\end{equation}
Note that it is assumed the chemical is produced at the tumour boundary at a constant rate, generating a source boundary condition for $c$.

\smallskip
\cite{owen1997} model for tumour-macrophage interactions, for macrophages ($l$), mutant cells ($m$), normal cells ($n$), chemical regulator ($f$) and macrophage-mutant complex ($c$):
\begin{equation}\tag{D12}\label{D12}
\begin{array}{rcl}
l_t &= & \nabla \cdot \left[ D_l \nabla l - \chi l \nabla f \right] + \frac{k_1 f l (k_2+k_3)}{k_4+l + m + n} - k_5(1+k_6 f) - k_7 f l m +k_8 e -k_9 l\,, \\
m_t & = & D_m \nabla^2 m + \frac{k_{10} f l (k_2+k_3)}{k_4+l + m + n} - k_{11} m - k_{7} f l m\,, \\
n_t & = & D_n \nabla^2 n + \frac{k_{11} f l (k_2+k_3)}{k_4+l + m + n} - k_{11} n\,, \\
f_t & = & D_f \nabla^2 f + k_{12} m - k_{13} f\,, \\
c_t & = & D_c \nabla^2 c + k_{7} f l m - k_{14} c - k_{15} c\,.
\end{array}
\end{equation}

\smallskip
\cite{knutsdottir2014} model for autocrine/paracrine macrophage-tumour interactions, for macrophages ($m$), tumour cells ($n$), CSF-1 concentration ($c$) and EGF concentration ($e$):
\begin{equation}\tag{D13}\label{D13}
\begin{array}{rcl}
m_t & = & \nabla \cdot \left[ D_m \nabla m - \chi_1 m \nabla c \right] \,,\\
n_t & = & \nabla \cdot \left[ D_n \nabla n - \chi_2 n \nabla e - \chi_3 n \nabla c \right] \,, \\
c_t & = & D_c \nabla^2 c + k_1 n - k_2 c \,,\\
e_t & = & D_e \nabla^2 e + k_3 m - k_4 e \,.
\end{array}
\end{equation}

\smallskip
\cite{chaplain2006} model for uPA mediated tumour invasion, for tumour cells ($c$), ECM ($v$) and uPA ($u$):
\begin{equation}\tag{D14}\label{D14}
\begin{array}{rcl}
c_t & = & \nabla \cdot \left[ D_c \nabla c - \chi_1 c \nabla u - \chi_2 c \nabla v \right] + k_1 c(1 - k_2 c - k_3 v) \,,\\
v_t & = & -k_4 u v +k_5 v (1 - k_2 c - k_3 v)\,, \\
u_t &= & D_u \nabla^2 u + k_6 c - k_7 u\,.
\end{array}
\end{equation}

\subsection{Models in ecology}

\cite{lee2009} model for predator-prey taxis, for predator ($p$) and prey ($q$):
\begin{equation}\tag{E1}\label{E1}
\begin{array}{rcl}
p_t & = & \nabla \cdot \left[ D_p \nabla p - \chi(q) p \nabla q \right] + k_1 p (F(p,q) - G(p))\,, \\
q_t &= & D_q \nabla^2 q + q H(q) - p F(p,q) \,,
\end{array}
\end{equation}
where $F, G, H$ are typical functions to describe predator-prey interactions.

\smallskip
\cite{tania2012} model for forager-scrounger interactions, for forager ($f$), scrounger ($s$) and food resource ($r$):
\begin{equation}\tag{E2}\label{E2}
\begin{array}{rcl}
f_t & = & \nabla \cdot \left[ D_f \nabla f - \chi_f f \nabla r \right] + G(f,s,r)\,;\\
s_t & = & \nabla \cdot \left[ D_s \nabla s - \chi_s s \nabla f \right] + H(f,s,r)\,;\\
r_t & = & D_r \nabla^2 r + k_1 - k_2 (f+s) r - k_3 r \,. 
\end{array}
\end{equation}
Forager/scrounger kinetics $G$ and $H$ could be zero, follow standard population growth 
terms or be chosen to describe within species strategy switching.

\smallskip
\cite{pearce2007} model for host-parasitoid chemotaxis systems, for two hosts ($n,m$), two parasites ($p,q$) and chemical ($c$):
\begin{equation}\tag{E3}\label{E3}
\begin{array}{rcl}
n_t & = & D_n \nabla^2 n +k_1 n (1-n/k_2) - k_3 p \left(1-e^{-k_4 n}\right)\,, \\
m_t & = & D_m \nabla^2 m +k_5 n (1-n/k_6) - k_7 p \left(1-e^{-k_8 m}\right) - k_9 q \left(1-e^{-k_{10} m}\right)\,, \\
p_t & = & \nabla \cdot \left[ D_p \nabla p - \chi_1 p \nabla c \right] + k_{11} p \left(1-e^{-k_4 n}\right) + k_{12} p \left(1-e^{-k_8 m}\right) - k_{13} p \,,\\
q_t & = & \nabla \cdot \left[ D_q \nabla q - \chi_2 q \nabla c \right]  + k_{14} q \left(1-e^{-k_{10} m}\right) - k_{15} q \,,\\
c_t & = & D_c \nabla^2 c + k_{16} (n+m)  - k_{17} c\,.
\end{array}
\end{equation}

\smallskip
\cite{ramakrishnan2014} model for ant foraging behaviour, for foraging ants ($u$), ants returning from food source $i$ ($v_i$) and pheromone ($c$):
\begin{equation}\tag{E4}\label{E4}
\begin{array}{rcl}
u_t & = & \nabla \cdot \left[ \nabla u - u \nabla c \right] + k_1 F(x) \sum_{i=1}^n v_i - \sum_{i=1}^{n} \kappa_i G_i(x) u\,,\\
{v_i}_t & = & - \nabla \cdot v_i \nabla F  + \kappa_i G_i(x) u - k_1 F(x) v_i \,,\\
c_t & = & D_c \nabla^2 c - c + \sum_{i=1}^{n} p_i u\,.
\end{array}
\end{equation}
Note that $F$ is a vector denoting the nest direction.

\smallskip
\cite{amorim2015} model for ant foraging behaviour, for foraging ants ($u$), returning ants ($w$), pheromone ($v$) and food resource ($c$):
\begin{equation}\tag{E5}\label{E5}
\begin{array}{rcl}
u_t & = & \nabla \cdot \left[ D_u \nabla u - \chi_u u \nabla v\right] -k_1 u c + k_2 w F(x) + G(t) F(x)\,, \\
w_t & = & \nabla \cdot \left[ D_w \nabla w - \chi_w w \nabla A\right] + k_1 u c - k_2 w F(x)\,, \\
v_t & = & D_v \nabla^2 v + k_3 H(x) w - k_4 v\,;\\
c_t & = & - k_5 u c\,.
\end{array}
\end{equation}
Note that $F(x)$ describes the location of the nest, $G(t)$ describes the rate of foraging ants emerging from the rest, $H(x)$ describes a decrease in pheromone deposition close to the nest and $\nabla A$ is an attraction to the nest.

\smallskip
\cite{logan1998} model for mountain pine beetle outbreaks, for flying beetles ($b$), nesting beetles ($n$), aggregation pheromone ($a$), host kairomones ($c$), tree resistance ($r$), and attack holes ($h$):
\begin{equation}\tag{E6}\label{E6}
\begin{array}{rcl}
b_t & = & \nabla \cdot \left[ D_b \nabla b - \chi_a b \nabla F(a) - \chi_c b \nabla c \right] + k_1 - k_2 b - k_3 \frac{r}{k_4} b \left(1+k_5 a \right)\,, \\
n_t & = & k_3 \frac{r}{k_4} b \left(1+k_5 a \right) - k_6 n - k_7 n r\,, \\
a_t & = & D_a \nabla^2 a + k_8 n - k_9 a\,,\\
c_t & = & D_c \nabla^2 c + k_{10} h r - k_{11} c\,,\\
r_t & = & r \left( k_{12} \left(k_{13} - r \right) - k_{14} h \right) \,, \\
h_t & = & k_3 \frac{r}{k_4} b \left(1+k_5 a \right) - k_{15} h r\,,
\end{array}
\end{equation}
where $\nabla F(a) = \frac{k_{16} - a}{k_{16} + a/k_{17}} \nabla a$.

\smallskip
\cite{strohm2013} model for mountain pine beetle outbreaks, for flying beetles ($b$), nesting beetles ($n$) and aggregation pheromone ($a$):
\begin{equation}\tag{E6}\label{E7}
\begin{array}{rcl}
b_t & = & \nabla \cdot \left[ D_b \nabla b - \chi b \frac{k_{1} - a}{k_{1} + a/k_{2}} \nabla a \right] + k_3 - k_4 b - k_5 b \frac{b^2}{b^2+k_6^2}\,, \\
n_t & = & k_5 b \frac{b^2}{b^2+k_6^2} - k_7 q\,, \\
a_t & = & D_a \nabla^2 a + k_8 n - k_9 a\,.
\end{array}
\end{equation}

\subsection{Models in sociology}

\smallskip
\cite{neto2015} model for capital-labour dynamics, for capital ($c$) and labour ($l$):
\begin{equation}\tag{F1}\label{F1}
\begin{array}{rcl}
l_t & = & \nabla \cdot \left[ D_l \nabla l -  \chi l \nabla c \right] + k_1 l (1-l/k_2)\,, \\
c_t & = & D_c \nabla^2 c + k_3 c^q l^{1-q} - k_4 c \,.
\end{array}
\end{equation}

\smallskip
\cite{short2008} model for crime hotspot formation, for criminals ($c$) and attractivity ($a,b$):
\begin{equation}\tag{F2}\label{F2}
\begin{array}{rcl}
c_t & = & D_c \nabla \cdot \left[ \nabla c - \frac{2 c}{b+a} \nabla (b+a) \right] + k_1 - c (b+a)\,, \\
a_t & = & D_a \nabla^2 a + k_2 c (b+a) - k_3 a \,.
\end{array}
\end{equation}
Note that the attractivity is decomposed into background level ($b(\bx)$) and  the dynamic component ($a$). 

\smallskip
\cite{pitcher2010} model for crime hotspot formation, for criminals ($c$), attractivity ($a,b$) and deterrent ($d$, policing):
\begin{equation}\tag{F3}\label{F3}
\begin{array}{rcl}
c_t & = & D_c \nabla \cdot \left[ \nabla c - \frac{2 c}{a(1-d)^+} \nabla a(1-d)^+ \right] + k_1 - k_2 c\,, \\
a_t & = & D_a \nabla^2 (a-b) + k_3 c a (1-d)^+ \left( 1-a/k_4 \right) - k_5 (a-b) \,, \\
d_t & = & D_d \nabla^2 d + k_6 u(\bx,t) - k_7 d
\end{array}
\end{equation}
where $u(t,\bx)$ describes deterrence, subject to resource constraint $\int_{\Omega} u d\bx = k_8$.

\medskip\noindent
{\bf References}
\bibliography{historyofchemotaxis}

\end{document}